\documentclass[aps,pre,a4paper,twocolumn,amsmath,showpacs]{revtex4}
\usepackage{graphicx}

\begin{document}
\bibliographystyle{apsrev}

\title{Surface phase transitions in polydisperse hard rod fluids}

\author{Yuri Mart\'{\i}nez-Rat\'on}
\email{yuri@math.uc3m.es} 
\affiliation{Grupo Interdisciplinar de Sistemas Complicados (GISC), \\ 
Departamento de Matem\'aticas, Universidad Carlos III de Madrid, \\
Avda.~de la Universidad 30, E--28911, Legan\'es, Madrid, Spain.} 

\date{\today}

\begin{abstract}
With the simple Zwanzig model in  Onsager 
approximation 
I study the effect of length polydispersity in the surface phase diagram  
of hard rods interacting with a hard wall. The properly extended interface 
Gibbs-Duhem equation for a polydisperse system allows us to predict the 
behaviour of the surface tension as a function of the bulk density at the 
the wall-isotropic interface. Two groups of qualitative different bulk and
surface phase diagrams are calculated from two families of parametrized  
length distribution functions $p(l)$. This parameterization controls the 
law of decay at large $l$. I also study the segregation due to polydispersity 
at the isotropic-nematic interface and the capillary nematization phenomena
as a function of polydispersity. 
\end{abstract}


\maketitle

\section{Introduction}
A molecular mixture with a finite number of components 
with different physical properties like characteristic dimension, mass or
charge is a well controlled system in the sense that all these properties 
are known exactly through the study of each component in its pure state. 
Also the relative proportion of each component 
is a parameter that can be controlled in experiments.  
The situation is very different in the industrial colloidal systems 
because during its manufacture process 
are involved many no controlled factors at microscopic scales. In 
this case these physical properties vary in a certain range practically 
in a continuous way. Then, is more appropriate to treat these system as 
polydisperse in nature with its properties distributed in some range 
according to some continuous size distribution function. 
Although in experiments are known that many 
systems are polydisperse in nature there is not  
information about the kind of 
polydisperse distribution 
function that follows its physical properties. Example of such systems 
are colloidal suspensions, polymer blends and emulsions.  
A common phenomena that have been observed 
experimentally in many polydisperse systems is that the coexisting phases 
have different particles sizes distributions \cite{Evans} 
and the simulations also corroborate these results for hard spheres \cite{Bolhuis}.
For anisotropic particles experiments shows that the polydispersity 
makes the phase diagram much more complex including up to four different coexisting 
phases \cite{Koij}. Recent experiments show that the isotropic 
phase of polydisperse hard disk system under certain 
conditions can be more dense than the nematic phase \cite{Beek}. 

From theoretical point of view the 
polydisperse systems composed by anisotropic particles have been in general 
modeled using a free energy functional for 
a binary mixture 
when the polydispersity has a clear bimodal distribution. 
The fundamental approach is then the mapping of polydisperse 
system onto a bidisperse one.These bidisperse systems 
show similar transitions found in experiments with the organophilic
polydisperse rod-like colloidal particles (boehmite core-particles)
including a three phase isotropic-nematic-nematic  region \cite{Lekk}.
Nevertheless the inclusion of bidispersity impose two relevant scales 
(the two characteristic lengths of the particles) that are absent in the 
unimodal polydisperse systems. 

Recent works include the polydispersity  
via a continuous size distribution function \cite{Speranza}. 
This was motivated by the development of the moment theory 
for polydisperse systems which makes more 
accessible the theoretical study because the original infinite number of 
variables that describe the whole system are reduced in this approximation 
to some moments 
of the distribution function through a projection to the space extended by them 
\cite{Sollich}. This approximation was applied to the study of melting 
in polydisperse hard spheres \cite{Bartlett}. Then, any density 
functional theory 
which excess part of free energy depends only on a few moments 
have a natural extension to the polydisperse case. The fundamental measure 
theory \cite{Rosenfeld}
is a clear candidate for applying it to these 
systems. This functional was used  in the study of the local size 
segregation of hard 
sphere fluid  in a presence of a hard wall \cite{Pagona}.

Although experimentally was obtained the bulk phase diagram for a few 
polydisperse systems including also the solid phases \cite{Pusey}
there is no so many works about inhomogeneous polydisperse 
systems where the inhomogeneities 
are created by external fields such as a hard wall or about the study  
of interfaces between the coexisting phases (there is a recent work \cite{Baus} 
about the study of the surface tension and adsorption properties of 
the planar interface between 
coexisting fluid phases).

Recently have been published some works about the study of hard rod 
monodisperse system in a presence of a hard wall or confined in the slit showing 
a continuous uniaxial-biaxial nematic transition at density bellow 
the bulk isotropic-nematic transition followed by the complete wetting 
by nematic at the wall-isotropic interface \cite{Roij}. Also was 
studied in these works the isotropic-nematic interface 
and the capillary nematization of the slit 
using the Zwanzig model in the Onsager approximation to describe the hard rod 
system. The simulations \cite{Roij} and experiments \cite{Kocevar} 
confirm in the qualitative level 
all these results.

In general, the calculation of inhomogeneous density profiles 
for mixtures
of hard core anisotropic bodies using density functional theory is 
a difficult problem for two reasons. One is the numerical problem  
caused by the presence of orientational degrees of 
freedom for each component, besides the spatial one. This  
complicates functional minimization. The second problem is that 
there are not so well developed and tested density functionals for mixtures as 
for one-component systems. Fundamental measure functionals (FMF) are 
good candidates because of its natural extension to mixtures from 
one-component systems 
\cite{Rosenfeld}. But these functionals 
which are very good at predicting the freezing of a hard spheres 
\cite{Ros-Schm} or the  
inhomogeneous density profiles induced by external 
fields \cite{Somoza}  have some difficulties if we try to extend them in a 
classical way to a system of  
anisotropic particles with 
continuous rotational degrees of freedom \cite{Chamoux}. The main problem here 
is that the Fourier transform of the 
direct correlation function calculated from the natural extension of 
the FMF to anisotropic particles is isotropic at $k=0$, which is 
unphysical \cite{Chamoux}. Nevertheless there is another way to construct 
a fundamental measure theory (FMT) 
without using the starting point of the scale particle 
theory. This is the cavities method \cite{Tarazona1}. This method 
refuses to postulate that the functional should depend only on 
certain weighted densities which are linear convolutions of the 
density profiles. The main idea behind it is to ensure the 
dimensional cross-over 
of the functional for different configurations of cavities \cite{Tarazona1}. 
The final result is a functional which depends on a nonlinear kernel that 
can not be decoupled in a set of one particle weighed densities. 
But this functional 
is impractical when there are rotational degrees of freedom, because the 
numerical minimization problem turns out to bee extremely hard. 

A simplification of the initial problem is to assume
that particles are 
restricted to have a few discrete orientations. Particles 
with different orientations are treated as different species and then 
it is completely justified to use the classical version of FMT 
for mixtures. The easiest way to implement this method 
is the Zwanzig model \cite{Zwanzig}, 
which describes a system of hard parallelepipeds 
with three mutual perpendicular orientations. An FMF for 
parallel hard parallelepipeds which fulfills all dimensional cross-overs \cite{Cuesta}  
has been applied to the study of the nature of the order of 
nematic-smectic A transition \cite{M-R}. 
I will use this functional 
in the Onsager limit  to study the continuous polydisperse system.
In Ref. \cite{Clarke} it has been presented the general formalism 
for the study of the bulk phase diagram of the 
polydisperse  
Zwanzig model in Onsager's limit. The bulk phase diagram was obtained 
for the particular 
choice of the Schulz size distribution function. 

I have two purposes in the study of the polydisperse hard rod system.
The first is that as I have already 
mentioned the knowledge of the specific type of polydispersity inherent to a 
particular system is commonly very poor. Then is important to study how 
depends the phase diagram topology on the kind of introduced polydispersity. 
After have been obtained as general as possible a characterization of the 
phase behaviour for different families of polydispersity distribution 
functions (I will suppose that all distributions are unimodal) 
I will study different interfaces: the wall isotropic, 
the isotropic-nematic and the nematic polydisperse hard rod system 
confined between two walls. An interesting questions are how polydispersity 
affect the surface phase transitions found in \cite{Roij} and to study  
the novel surface behaviour inherent only to polydisperse systems. 

This paper is organized as follows. In section \ref{model} is obtained 
formally the Onsager limit of the Fundamental Measure Functional 
for polydisperse 
hard parallelepipeds and the proper equations to calculate the bulk 
phase diagram (subsection \ref{bulkphase}) and the 
different interfaces (subsection \ref{interface}). 
In section \ref{thermo} are deduced 
some general results for thermodynamic of polydisperse interfaces, 
specifically the extension of the interface 
Gibbs-Duhem equation to polydisperse 
case. In section \ref{finite} are presented some important remarks 
about the validity of the thermodynamic limit in finite polydisperse 
systems. The results are presented 
in section \ref{results} divided in different subsections 
where I study the possible bulk and surface phase diagrams for 
rapidly (subsection \ref{rapidly}) and slowly (subsection \ref{slowly}) 
decaying distribution 
functions.   
The last subsections are devoted to the study of 
different interfaces: wall-isotropic, isotropic-nematic 
(subsection \ref{w-i-n}) and a polydisperse 
nematic liquid crystal in the slit (subsection \ref{capillary}). 
Finally the conclusions are summarized in section 
\ref{conclusions}.

\section{Model}
\label{model}
\begin{figure}
\mbox{\includegraphics*[width=2.5in, angle=0]{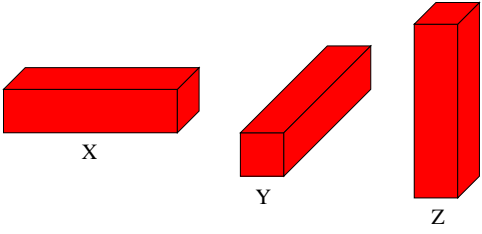}}
\caption[]{Zwanzig model: the uniaxial axes of parallelepipeds 
are oriented to one of the three 
perpendicular directions.}
\label{fig1}
\end{figure}

Suppose we have a multicomponent mixture of uniaxial parallelepipeds, 
all of them with the same cross section width $\sigma$ but different lengths 
$\{L_{\nu}\}$ ($\nu=1,\dots c$, where $c$ is the number of species). 
The orientation 
of each parallelepiped is restricted to one of the three 
perpendicular axis $x$, $y$ or 
$z$ as is sketched in figure \ref{fig1}. 
This is the well known Zwanzig model \cite{Zwanzig}. 
As I want to study planar interfaces I will suppose that the density 
profiles of different species are 
inhomogeneous only in one spatial direction and I will label them
as follows: 
$\rho_{\mu}(L_{\nu},z)$, where $\mu=x,y,z$ 
labels the orientation of the species with length $L_{\nu}$. 

The excess part of free energy density 
according to the fundamental measure theory for this model is 
\cite{Cuesta}  
\begin{eqnarray}
\Phi_{ex}=-n_0\ln(1-n_3)+\frac{{\bf n}_1\cdot{\bf n}_2}{1-n_3}+
\frac{ n_{2x} n_{2y} n_{2z}}{(1-n_3)^2},
\label{eq1}
\end{eqnarray}
where the $n_{\alpha}$'s are weighted densities, i.e.
linear convolutions of the density profiles with certain weights:
\begin{eqnarray}
n_{\alpha}=\sum_{\nu=1}^c\sum_{\mu={x,y,z}}\rho_{\mu}(L_{\nu},z)
\ast w^{(\alpha)}_{\mu}(L_{\nu},z), 
\label{eq2}
\end{eqnarray}
where ``$\ast$'' means the convolution with respect to the spatial variables. 
These weights are characteristic 
functions whose integrals are fundamental measures of the convex 
particle \cite{Rosenfeld,Cuesta}. 
The nature of the $w^{(\alpha)}$'s can be either 
scalar or vector. The scalars are 
\begin{eqnarray}
\omega^{(0)}_{\mu}(L_{\nu})=\delta^{\mu}_x(L_{\nu})\delta^{\mu}_y(L_{\nu})
\delta^{\mu}_z(L_{\nu}), \label{eq3a}\\
\omega^{(3)}_{\mu}(L_{\nu})=\theta^{\mu}_x(L_{\nu})\theta^{\mu}_y(L_{\nu})
\theta^{\mu}_z(L_{\nu}), \label{eq3b}
\end{eqnarray}
where for example $\delta^{z}_z(L_{\nu})=\delta(L_{\nu}/2-|z|)$ and 
$\delta^y_z(L_{\nu})=\delta(\sigma/2-|z|)$. These examples allow us to 
recognize the used notation: the subindex labels the spatial coordinate present 
in the 
argument of the Dirac delta function. If the superindex and the subindex 
coincide then the corresponding parallelepiped edge length in the argument 
of this function is $L_{\nu}$ otherwise    
is $\sigma$. The meaning of the labels for 
$\theta^{\mu}_{\sigma}(L_{\nu})$ 
is the same as in the preceding case, but $\theta$ now represents     
Hevisaide function $\Theta(x)$.

The vector weighs are 

\begin{eqnarray}
{\bf w}_{\mu}^{(i)}(L_{\nu})=\left(w^{(i)}_{\mu x}(L_{\nu}), 
w^{(i)}_{\mu y}(L_{\nu}), w^{(i)}_{\mu z}(L_{\nu})\right), 
\label{eq4}
\end{eqnarray}
with $i=1,2$ and
\begin{eqnarray}
w^{(1)}_{\mu x}(L_{\nu})&=&\theta^{\mu}_x (L_{\nu})\delta^{\mu }_y(L_{\nu})
\delta^{\mu }_z(L_{\nu}), \\
w^{(2)}_{\mu x}(L_{\nu})&=&\delta^{\mu }_x(L_{\nu})\theta^{\mu }_y(L_{\nu})
\theta^{\mu }_z(L_{\nu}).
\end{eqnarray}
The component $w^{(1)}_{\mu x}(L_{\nu})$ is obtained from $\omega^{(0)}_{\mu} 
(L_{\nu})$  
substituting in Eq. (\ref{eq3a}) $\delta^{\mu}_x(L_{\nu})$ 
by $\theta^{\mu}_x(L_{\nu})$
and  $w^{(2)}_{\mu x}(L_{\nu})$ is obtained from $\omega^{(3)}_{\mu}(L_{\nu})$ 
(see Eq. (\ref{eq3b}))
doing the opposite substitution. 
The rest of the components can be 
obtained using the same procedure.

Now, suppose that we select a natural scale $L_0$ for the 
length of the parallelepipeds, 
for example the mean length $L_0=\sum_{\mu} L_{\mu}x_{\mu}$ (where 
$x_{\mu}$ is the molar fraction of the specie ${\mu}$). 
Doing the change of variable $z\to z/L_0$ and 
$l_{\nu}=L_{\nu}/L_0$ and taking the limit  $c\to \infty$ (infinite 
number of components, so the mixture contains parallelepipeds with 
lengths distributed with certain continuous function) 
we should replace $\sum_{\nu}\to \int dl$ in any 
of the equations that define the 
weighted densities, and the density profiles  
are now functions of $l$ and $z$: $\rho(l,z)$. We should remember that $z$ 
is in reduced units.

We know from Onsager that the correct  variable to describe a nematic 
phase in the limit of infinite aspect ratio is $\rho L_0^2\sigma$, so 
I will use in the following this scaled density for any specie using 
the same notation: $\rho_{\mu}(l,z)L_0^2\sigma\to\rho_{\mu}(l,z)$.
Using the expansions  
of the convolutions $\rho_{\mu}(l,z)*\delta_z^{\mu}$ and 
$\rho_{\mu}(l,z)*\theta_z^{\mu}$ ($\mu=x,y$) in powers of $\sigma$:
\begin{eqnarray}
\rho_{\mu}(l,z)+\frac{\rho''_{\mu}(l,z)}{8}\sigma^2+\cdots \\
\text{and} \quad \rho_{\mu}(l,z)\sigma+\frac{\rho''_{\mu}(l,z)}{24}\sigma^3+\cdots
\end{eqnarray}
respectively (primes denote derivatives w.r.t. $z$) and expanding 
the excess part of the free energy density (\ref{eq1}) 
in powers of $\lambda^{-1}=\sigma/L_0$ 
is obtained
\begin{eqnarray}
\Phi_{ex}L_0^2\sigma=\Phi_0+\lambda^{-1}\Phi_1+\cdots,
\end{eqnarray}
with
\begin{eqnarray}
\Phi_0&=&\sum_{\mu}m_{\mu}\Psi_{\mu},\\
\Phi_1&=&m(\rho+\Phi_0)+\sum_{\mu}\Psi_{\mu}\Upsilon_{\mu}+
\prod_{\mu=x,y,z}\Upsilon_{\mu}.
\end{eqnarray}
where $\rho=\sum_{\mu}\rho_{\mu}$  and $m=\sum_{\mu}m_{\mu}$ (was dropped the 
dependence of $\rho_{\mu}$ and $m_{\mu}$ on $z$ for notation simplicity).
I have used the following definitions
\begin{eqnarray}
\Psi_{\mu}=\sum_{\nu \neq \mu}m_{\nu}, \quad
\Upsilon_{\mu}=\sum_{\nu\neq \mu} \rho_{\nu}, \quad \mu,\nu=x,y,z
\end{eqnarray}
where
\begin{eqnarray}
\rho_{\mu}(z)&=&\int dl \rho_{\mu}(l,z), 
\quad \rho_z(z)=\int dl \rho_z(l,z)\ast\delta_l, \label{eq5a}\\
m_{\mu}(z)&=&\int dl l \rho_{\mu}(l,z) , \quad 
m_z(z)=\int dl \rho_z(l,z)\ast \theta_l, \nonumber\\
\label{eq5b}
\end{eqnarray}
with $\mu=x,y$, $\delta_l=\delta(l/2-|z|)$ and $\theta_l=\Theta(l/2-|z|)$.
$\Phi_0$ is then the Onsager limit  
whereas $\Phi_1$ is the first correction. It is interesting to notice that
$\Phi_0$ depends only on the first moments of $\rho_{\mu}(l,z)$ while
$\Phi_1$ depends on the zeroth moments as well.

Then, our approximation for the   
total free energy per unit area when the system is 
inhomogeneous in one direction is just
\begin{eqnarray}
\beta{\cal F}L_0\sigma/A=\int dz\left(\Phi_{id}(z)+\Phi_0(z)\right), 
\label{eq6}
\end{eqnarray}
where was done the proper scaling of the ideal part 
$\Phi_{id}\to\Phi_{id}L_0^2\sigma$ which has the form
\begin{eqnarray}
\Phi_{id}&=&\sum_{\mu}\int dl \rho_{\mu}(l,z)\left[
\ln \left(\rho_{\mu}(l,z)\Lambda_{\mu}^3(l)\right)-1\right], \label{eq7} 
\end{eqnarray}
with 
$\Lambda_{\mu}^3(l)$ the cube of the thermal lengths of specie 
$\mu$.  

\subsection{Bulk phases}
\label{bulkphase}
The bulk phase diagram is calculated using 
the obvious homogeneities condition $\rho_{\mu}(l,z)=\rho_{\mu}(l)$. 
The total density of the rods with length $l$ is 
$\rho(l)=\sum_{\mu}\rho_{\mu}(l)$, and the mean density of the system 
$\rho=\int dl \rho(l)$. The fraction of rods with length $l$ 
parallel to the $\mu$ axis (calculated from 
$P_{\mu}(l)\equiv\rho_{\mu}(l)/\rho(l)$) is a measure 
of the orientational order and so the functional $\Phi=\Phi_{id}+\Phi_0$ 
should be 
minimized with respect to it with the obvious constraint 
$\sum_{\mu}P_{\mu}(l)=1$. After this functional minimization is obtained
\begin{eqnarray}
P_{\mu}(l)=\frac{e^{-2\Psi_{\mu}l}}{\sum_{\nu}e^{-2\Psi_{\nu}l}}.
\label{eq8}
\end{eqnarray}
As usual, the chemical potential of rods with length $l$ is calculated from 
$\displaystyle{\beta \mu(l)=\frac{\delta \Phi}{\delta \rho(l)}}$.
Using this definition and (\ref{eq8}) is obtained the result
\begin{eqnarray}
\beta \mu(l)=\ln\left(\frac{\rho(l)}{\sum_{\mu}e^{-2\Psi_{\mu}l}}\right).
\label{eq9}
\end{eqnarray}
Finally, the reduced osmotic pressure of the system is 
calculated from the thermodynamic 
relation: $\beta \Pi=\int dl\rho(l)\beta \mu(l)-\Phi$, 
with the following result
\begin{eqnarray}
\beta \Pi=\rho+\Phi_0.
\end{eqnarray}
The coexistence between $q$ phases is calculated from the equalities 
between the chemical potentials and pressures of each type of 
rod for the $q$ different phases.  
Selecting a reference 
isotropic phase (referred to as the parent phase and labelled with 
superindex $0$) with some fixed 
probability length distribution function then the set of coexisting equations 
are
\begin{eqnarray}
\mu^{(\alpha)}(l)&=&\mu^{(0)}(l),\\
\Pi^{(\alpha)}&=&\Pi^{(0)},\quad \alpha=1,\dots,q. \label{eq10}
\end{eqnarray}
The first set of equations with (\ref{eq9}) give us the 
expressions for the densities of each phase
\begin{eqnarray}
\rho^{(\alpha)}(l)=e^{\beta \mu^{(0)}(l)}\sum_{\nu}e^{-2\Psi_{\nu}^{(\alpha)}l}.
\label{eq11}
\end{eqnarray}
If we have an initial polydisperse parent phase with density 
distribution $\rho^{(0)}(l)$ then the $q$ coexisting densities  
$\rho^{(\alpha)}(l)$ should obey the conservation law 
\begin{eqnarray}
\sum_{\alpha}x_{\alpha}\rho^{(\alpha)}(l)=\rho^{(0)}(l), 
\label{eq11a}
\end{eqnarray}
where $x_{\alpha}$ is the volume fraction of the $\alpha$ phase 
($\sum_{\alpha}x_{\alpha}=1$). 

Using (\ref{eq11a}) we obtain from (\ref{eq9})and (\ref{eq11}):
\begin{eqnarray}
\rho^{(\alpha)}(l)=\frac{\rho^{(0)}(l)\sum_{\mu} e^{-2\Psi_{\mu}^{(\alpha)}l}}
{\sum_{\beta}x_{\beta}\sum_{\nu}e^{-2\Psi^{(\beta)}_{\nu}l}},
\end{eqnarray}
and using the definition $\rho^{(\alpha)}_{\mu}(l)=P_{\mu}^{(\alpha)}(l)
\rho^{(\alpha)}(l)$ we have 
\begin{eqnarray}
\rho_{\mu}^{(\alpha)}(l)=\frac{\rho^{(0)}(l)e^{-2\Psi_{\mu}^{(\alpha)}l}}
{\sum_{\beta}x_{\beta}\sum_{\nu}e^{-2\Psi_{\nu}^{(\beta)}l}}.
\label{eq12}
\end{eqnarray}
The definitions of the first moments, Eq. (\ref{eq5b})  
with $\rho_z(l,z)\ast\theta_l
=\rho_z(l)l$, and Eq. (\ref{eq12}) 
allow us to derive a self consistent set 
of $3q$ equations 
\begin{eqnarray}
m_{\mu}^{(\alpha)}=\int dl l\frac{\rho^{(0)}(l)e^{-2\Psi^{(\alpha)}_{\mu}l}}
{\sum_{\beta}x_{\beta}\sum_{\nu}e^{-2\Psi_{\nu}^{(\beta)}l}}, \quad 
\mu=x,y,z, 
\label{eq13}
\end{eqnarray}
with $\alpha=1,\dots q$.
If $\rho_0$ is 
the mean density of the parent phase $\rho_0\equiv \int dl \rho^{(0)}(l)$ 
with $p(l)$ its length distribution function, 
$\rho(l)\equiv\rho_0p(l)$, we 
have $4q$ unknowns: the $3q$ moments $m_{\mu}^{(\alpha)}$, 
the $q-1$ independent $x_{\alpha}$ 
and $\rho_0$, and $4q$ equations 
to solve: 
the set (\ref{eq13}) of $3q$ equations and the set of $q$ 
equations for the equalities of pressures. 

Now I will specify the model for a two phase coexistence 
between an isotropic and a uniaxial nematic phase. For the nematic 
phase we have $m_y^{(N)}=m_z^{(N)}\equiv m_{\perp}$ and $m_x^{(N)}=m_{\parallel}^{(N)}$, 
and for the isotropic one $m_x^{(I)}=m_y^{(I)}=m_z^{(I)}
\equiv \frac{1}{3}m^{(I)}$, where $m^{(I)}$ is the total moment of the 
isotropic phase. 
Defining the new variables  
\begin{eqnarray}
\tau=m_{\parallel}+m_{\perp}-\frac{2}{3}m^{(I)}, \quad s=m_{\parallel}-
m_{\perp},
\label{eq14}
\end{eqnarray}
and using the shorthand notation  
\begin{eqnarray}
E(l)=3x+(1-x)\epsilon(l), \quad 
\epsilon(l)=e^{-2\tau l}\left(e^{2sl}+2\right),
\label{eq15}
\end{eqnarray}
where $x$ is the fraction of volume occupied by the isotropic phase, 
we obtain from (\ref{eq13}) the equations for $\tau$ and $s$:
\begin{eqnarray}
s&=&\rho_0\phi_s, \quad
\phi_s=\int dl lp(l)\frac{e^{-2\tau l}\left(e^{2sl}-1\right)}{E(l)}, 
\label{eq16}\\
\tau&=&\rho_0\phi_{\tau}, \quad \phi_{\tau}=\int dl l p(l)
\frac{e^{-2\tau l}\left(e^{2sl}+1\right)-2}{E(l)}. \label{eq17}
\end{eqnarray}
The quantities $\rho^{(I)}$, $\rho^{(N)}$ (the total densities of 
the isotropic and nematic phases), and $m^{(I)}$, $m^{(N)}$ (the 
total first moments of the isotropic and nematic phases),
are functions of $\tau$ and $s$:
\begin{eqnarray}
\rho^{(I)}&=&\rho_0\phi_{\rho^{(I)}}, \quad \phi_{\rho^{(I)}}=
3\int dl p(l)\frac{1}{E(l)},  \label{eq18}\\
\rho^{(N)}&=&\rho_0\phi_{\rho^{(N)}}, \quad \phi_{\rho^{(N)}}=
\int dl p(l) \frac{\epsilon(l)}{E(l)}, \label{eq19}\\
m^{(I)}&=&\rho_0 \phi_{m^{(I)}}, \quad \phi_{m^{(I)}}=
3\int dl lp(l)\frac{1}{E(l)}, \label{eq20}\\
m^{(N)}&=&\rho_0\phi_{m^{(N)}}, \quad \phi_{m^{(N)}}=
\int dl l p(l)\frac{\epsilon(l)}{E(l)}. \label{eq21}
\end{eqnarray}

The pressures of the isotropic and nematic phases are related 
with the new variables $\tau$ and $s$ via 
\begin{eqnarray}
\beta\Pi^{(I)}&=&\rho^{(I)}+\frac{2}{3}{m^{(I)}}^2, \\
\beta \Delta\Pi&=&\Delta\rho+\frac{(\tau-s)(3\tau+s)}{2}+\frac{2m^{(I)}
(3\tau-s)}{3},\nonumber \\
\label{eq22}
\end{eqnarray}
where $\Delta \Pi=\Pi^{(N)}-\Pi^{(I)}$ and $\Delta \rho=\rho^{(N)}-\rho^{(I)}$.
To ensure the coexistence between the isotropic and nematic phases 
$\Delta \Pi$ should be equal to zero. Then, from (\ref{eq22}) and 
(\ref{eq18})-(\ref{eq21}) is possible to 
calculate $\rho_0$ as a function of $\tau$ and $s$:
\begin{eqnarray}
\rho_0=\frac{\phi_{\rho^{(N)}}-\phi_{\rho^{(I)}}}{
\frac{1}{2}(\phi_s-\phi_{\tau})(3\phi_{\tau}+\phi_s)-
\frac{2}{3}\phi_{m^{(I)}}(3\phi_{\tau}-\phi_s)}.
\label{eq23}
\end{eqnarray}
Finally, to find the coexistence between the isotropic and nematic 
phases we need to solve the closure set of two equations (\ref{eq16}) and 
(\ref{eq17}) (using expression (\ref{eq23})) 
with two unknowns $\tau$ and $s$. These are the equations used below 
to calculate numerically, for a fixed fraction of volume of the isotropic 
phase, $x$, the densities of the isotropic and nematic coexisting phases
as a function of the 
polydispersity. I will use the experimental nomenclature to classify 
the different coexisting curves. If there is an infinitesimal  
amount of nematic phase coexisting with an isotropic phase ($x=1$)
the dependence of the isotropic density with polydispersity is called 
the cloud isotropic curve, whereas the nematic density as function 
of polydispersity is called the shadow nematic curve. 
In the other extreme case where an infinitesimal amount of isotropic 
phase coexists with a nematic phase ($x=0$) we have  
the nematic cloud and the isotropic shadow curves.

\subsection{Interface}
\label{interface}

For the inhomogenous system in one spatial direction (the $z$ direction) 
was obtained (Eqs. (\ref{eq6}-\ref{eq7}) 
the Helmholtz density functional 
per unit area.  
 Then the grand potential is, by definition, 
 \begin{eqnarray}
 \frac{\Omega L_0\sigma}{A}&=&\frac{{\cal F}L_0\sigma}{A} \nonumber 
 \\
&+& \sum_{\mu}\int dz \int dl \rho_{\mu}(l,z)\left(
 v_{\mu}(l,z)-\mu(l)\right),
 \label{eq24}
 \end{eqnarray}
where $v_{\mu}(l,z)$ is the external potential acting on a rod of species 
$\mu$ and length $l$.  
A hard wall placed at $z=0$ can be described with the following 
external potential 
\begin{eqnarray}
v_z(l,z)=
\begin{cases}
\infty &  z< \frac{l}{2} \\
0  &  z \ge \frac{l}{2}
\end{cases}
\label{eq25}
\end{eqnarray}
and $v_{x}(l,z)=v_{y}(l,z)=0$ for all $z\ge 0$.
I will have assumed that at infinite distance from  
the wall the stable phase is the isotropic one, with 
mean density $\rho_0$, length distribution $p(l)$ and mean length fixed to 
unity. 
In  this case the chemical 
potential for each species is the same ( 
$\mu_x(l)=\mu_y(l)=\mu_z(l)=\mu^{(0)}(l)$). From (\ref{eq9}) we have  
\begin{eqnarray}
\beta \mu^{(0)}(l)= \ln\left(\frac{\rho_0}{3}p(l)e^{2l\Psi^{(0)}}\right).
\end{eqnarray}
Inserting (\ref{eq25}) in (\ref{eq24}) 
and minimizing $\Omega$ with respect to $\rho_{\mu}(l,z)$  
we obtain 
\begin{eqnarray}
\rho_{\mu}(l,z)&=&\frac{\rho_0}{3}p(l)e^{-2l\Delta \Psi_{\mu}(z)}, 
\quad \mu=x,y \label{eq26},\\
\rho_z(l,z)&=&\frac{\rho_0}{3}p(l)e^{-2\Delta \Psi_{z}\ast 
\theta_l(z)}\Theta(z-\frac{l}{2}),
\label{eq27}
\end{eqnarray}
with the definition $\Delta \Psi_{\mu}(z)=\Psi_{\mu}(z)-\Psi_{\mu}^{(0)}$ 
and $\Psi_{\mu}^{(0)}=2m_0/3$ where $m_0=m_{\mu}(\infty)$. 
A self-consistent 
set of equations for the first moments can be obtained from 
(\ref{eq26}), (\ref{eq27}) and (\ref{eq5b}), 
with the new variables (\ref{eq14}) (identifying $m_{\parallel}$ with 
$m_x$ and $m_{\perp}$ with $m_y$ which now depend on $z$) and  
\begin{eqnarray}
\vartheta(z)=
2(m_z(z)-m_0),
\end{eqnarray}
The result is (I drop the explicit 
dependence on $z$ for notational simplicity)
\begin{eqnarray}
\tau&=&\frac{2}{3}\rho_0\left(\int dl lp(l)e^{-(\tau+\vartheta)l}
\cosh(ls)-1\right), \label{eq28}\\
s&=&\frac{2}{3}\rho_0\int dl lp(l)e^{-(\tau+\vartheta)l}\sinh(sl), \label{eq29}\\
\vartheta&=&\frac{2}{3}\rho_0\left(\int dlp(l)\int_{\rm{max}(z-l/2,l/2)}
^{z+l/2}dz'e^{-2\tau\ast \theta_l(z')}-1\right). \nonumber \\ \label{eq30}
\end{eqnarray}
In terms of these variables the equilibrium density profiles, after 
finding the solution to (\ref{eq28})-(\ref{eq30}), have the form 
\begin{eqnarray}
\rho_x&=&\frac{\rho_0}{3}\int dl p(l)e^{-(\tau+\vartheta-s)l}, 
\label{eq31}\\
\rho_y&=&\frac{\rho_0}{3}\int dl p(l)e^{-(\tau+\vartheta+s)l}, 
\label{eq32}\\
\rho_z&=&\frac{\rho_0}{3}\int_0^{2z}dl p(l)e^{-2\tau\ast \theta_l}.
\label{eq33}
\end{eqnarray}
From  (\ref{eq31}) and (\ref{eq32}) we can  define 
an ``in plane'' order parameter as 
\begin{eqnarray}
Q_b=\frac{\rho_x-\rho_y}{\rho_x+\rho_y}=\frac{\int dl p(l)e^{-(\vartheta+
\tau)l}\sinh(sl)}{\int dl p(l)e^{-(\vartheta+\tau)l}\cosh(sl)}.
\label{eq34}
\end{eqnarray}
The model can be further simplify by a locality assumption:
is approximated any convolution $f\ast \theta_l(z)$ 
by $f(z)l$. Equations (\ref{eq28}) and (\ref{eq29}) then remain 
unchanged, but  
(\ref{eq30}) and (\ref{eq33}) change to 
\begin{eqnarray}
\vartheta&=&\frac{2}{3}\rho_0\left(\int_0^{2z}dllp(l)e^{-2\tau l}-1\right), 
\label{eq35}\\
\rho_z&=&\frac{\rho_0}{3}\int_0^{2z}dl p(l)e^{-2\tau l}.
\label{eq36}
\end{eqnarray}
The equations (\ref{eq28})-(\ref{eq29}), plus either (\ref{eq30}) or 
(\ref{eq35}), are our starting point 
for the study of the interface phase behaviour of 
the polydisperse hard rod system in the presence of a hard wall.  

In \cite{Roij1} it was studied, theoretically and by simulations, 
the wetting by nematic at the wall-isotropic 
interface for the one-component hard rod system, as well as 
the capillary nematization 
of the same system confined between two walls. For modeling the 
system the Zwanzig approximation in the Onsager limit was used.
The hard wall generates orientational order 
because is entropically favored the alignment of rods parallel to the wall.  
This orientational order enhanced by the wall translates  
into the  density profile in the creation of a nematic layer 
whose thickness increases continuously 
from zero to infinity when the bulk density is increased from 
$\rho_0^{\ast}$  
(the density at which the second order 
uniaxial-biaxial nematic transition occurs) to $\rho_w$ ( 
the complete wetting density) \cite{Roij1}.
An interesting question to study is the effect of polydispersity  
on the surface phase diagram. Of course this effect depends on 
the length distribution function of the isotropic 
bulk phase $p(l)$, but as I will show below, the  
surface phase behaviour for any $p(l)$ belongs
to one out of two qualitatively very different phase diagrams, which are 
determined by the behaviour of $p(l)$ at long lengths.

Whenever the uniaxial-biaxial nematic transition at the wall 
is a continuous transition we can calculate the 
transition density $\rho_0^{\ast}$ for any $p(l)$ as I will show below. 
The value of the total density and the first moment at contact 
can also be obtained analytically. 

At the wall, the value of $\vartheta(0)$ is $-2\rho_0/3$  
as can be easily seen from Eq. (\ref{eq30}). At 
infinity we have $\vartheta(\infty)=0$ because of the boundary conditions. 
Then we can solve the set of two equations 
(\ref{eq28}) and (\ref{eq29}) for any value of $\vartheta(z)$ 
in the interval $[-2\rho_0/3,0]$. 

Calling $\kappa=\tau+\vartheta$, the Eqs. (\ref{eq28})-(\ref{eq29})
can be put in a more convenient form to make these calculations: 
\begin{eqnarray}
s&=&\frac{2}{3}\rho_0\int dl l p(l)e^{-\kappa l}\sinh sl, \label{eq37} \\
\kappa&=&\vartheta+\frac{2}{3}\rho_0\int dl l p(l)\left(e^{-\kappa l}
\cosh sl-1\right)
\label{eq38}
\end{eqnarray}
The variable $s$ calculated from the equation $s=f(\kappa,s)$, where $f$ is 
the right hand side of Eq. (\ref{eq37}), is 
a kind of surface order parameter, which is 
zero below the transition density $\rho_0^{\ast}$ and is positive above it 
(I will assume that the symmetry breaking at the wall 
occurs in the $x$ direction). 

If the transition is continuous then $f_s'(\kappa^{\ast},0)=1$
should be obey at the transition point. Using this fact and  combining 
Eqs. (\ref{eq37}) and (\ref{eq38}), we obtain one single equation to calculate
the bifurcation value $\kappa^{\ast}$, 
\begin{eqnarray}
\int dl lp(l)e^{-\kappa^{\ast}l}\left(1-\kappa^{\ast} l\right)=2,
\label{eq39}
\end{eqnarray}
and from it the bifurcation density,  
\begin{eqnarray}
\rho_0^{\ast}=\frac{3}{2}\left[\int dl l^2 p(l)e^{-\kappa^{\ast}l}\right]^{-1}.
\label{eq40}
\end{eqnarray}
Equations (\ref{eq39}) and (\ref{eq40}) will be used below for the 
calculations of the second order uniaxial-biaxial nematic transition 
at the wall for 
different length distribution functions.

\section{Thermodynamics of polydisperse interfaces}
\label{thermo}
Once the grand potential is minimized we can calculate the 
surface tension 
\begin{eqnarray}
\gamma=\frac{\Omega}{A} +\Pi_0 L,
\label{eq41}
\end{eqnarray}
where $L$ is the length of interface and $\Pi_0$ the bulk osmotic pressure.
To characterize surface transitions (as wetting) is 
important to calculate  the density adsorption coefficient. For the 
polydisperse case I will 
show that, apart from it, the other moment adsorption coefficients   
of the total density distribution function are also relevant. In the  
excess part of free energy functional of this model 
only the first moment appears ,so the  
quantities we are referring to are
\begin{eqnarray}
\Gamma^{(0)}(\rho_0)&=&\sum_{\mu}\int_0^{\infty}dz\left(\rho_{\mu}(z)-
\frac{\rho_0}{3}\right), \label{eq42}\\
\Gamma^{(1)}(\rho_0)&=&\sum_{\mu}\int_0^{\infty}dz\left( m_{\mu}(z)-m_0\right).
\label{eq43}
\end{eqnarray}
In a mixture of a finite number of species the interface Gibbs-Duhem equation 
relates surface tension changes with changes of the chemical 
potentials of the different components as in
\begin{eqnarray}
0=d\gamma+\sum_{\nu}\Gamma_{\nu}d\mu_{\nu},
\label{eq44}
\end{eqnarray}
where $\Gamma_{\nu}$ is the adsorption coefficient of 
component $\nu$, and $\mu_{\nu}$ 
its chemical potential. 
The natural extension of Eq. (\ref{eq44}) to the polydisperse case is 
\begin{eqnarray}
d\gamma=-\int dl \Gamma(l) d\mu(l),
\label{eq45}
\end{eqnarray}
where $\Gamma(l)=\int_0^{\infty}dz\left(\rho(l,z)-\rho^{(0)}(l,\infty)\right)$.
Suppose the system is composed by a continuous mixture of asymmetric 
particles with one of its characteristic lengths, say   
$l$ (from the complete set of characteristic lengths that 
define the geometry of the particle), 
taking different values according to some fixed 
length distribution function $p(l)$, whereas the other lengths
are constant.
In general, the chemical potential of the species with length  
``$l$'' in the bulk 
isotropic phase with total density $\rho_0$ can be expressed, 
according to the scale 
particle theory, as 
\begin{eqnarray}
\mu(l)=\ln\left(\rho_0 p(l)\right)+\sum_{k=0}^{D'} a_k(\rho_0)l^k,
\label{eq46}
\end{eqnarray}
where the first term comes from the ideal part and the second one 
from the excess part of the free energy density, which in any scale particle 
theory   
is a polynomial on $l$ of degree $D'$ with coefficients depending on $\rho_0$.
$D'$ is a number between 1 and $D$ (the dimension of the 
system) which depends on the specific geometry of particles.    
For example, in three dimensions  we have $D'=3$ for 
hard spheres, $D'=2$ for uniaxial oblate parallelepipeds, or $D'=1$ for 
uniaxial prolate 
parallelepipeds.

A change in the bulk density $\rho_0$ implies a change in the surface 
tension, and using Eqs. (\ref{eq45}) and (\ref{eq46}) we 
can find an expression for the derivative of the surface tension with 
respect to the bulk density,
\begin{eqnarray}
\frac{d \gamma}{d\rho_0}=-\frac{\Gamma^{(0)}}{\rho_0}-
\sum_{k=0}^{D'}a_k'(\rho_0)\Gamma^{(k)},
\label{eq47}
\end{eqnarray}
where
\begin{eqnarray}
\Gamma^{(k)}=\int_0^{\infty}dz \left(M^{(k)}(z)-M^{(k)}(\infty)\right)
\end{eqnarray}
is the $k$-moment adsorption coefficient. The $k$-moment is defined as 
usual: $M^{(k)}(z)=\int dl l^k \rho(l,z)$.   

This important result tells us that the slope of the surface tension 
depends in general on different moment adsorption coefficients of the density 
distribution function $\rho(l,z)$. It is thus possible to find a 
qualitatively different behaviour of the
surface tension as compared to that of the one-component system, 
for which the slope depends only on the density adsorption coefficient, as 
we can see after substituting $\rho(l,z)=\rho(z)\delta(l_0-l)$ in 
(\ref{eq47}):   
\begin{eqnarray}
\frac{d\gamma}{d\rho_0}=-\left[\frac{1}{\rho_0}+\sum_{k=0}^{D'}
a_k'(\rho_0)l_0^k\right]\Gamma^{(0)}.
\label{eq48}
\end{eqnarray}

For example, as I will show below, fixing the degree of polydispersity 
we can find some range of $\rho_0$'s for which the density adsorption 
coefficient $\Gamma^{(0)}$ and the slope $d\gamma/d\rho_0$ are both negative, 
in contrast with the mono-disperse case behaviour (see Eq. (\ref{eq48})),  
for which $d\gamma/d\rho_0$ and $\Gamma^{(0)}$ have always opposite signs. This 
is due to the fact that the first moment adsorption coefficient $\Gamma^{(1)}$ 
is still positive as a consequence of the higher depletion effect that the 
wall induces on the large rods. Then, 
from Eq. (\ref{eq47}) we can see that the final sign depends 
on the modules of $\Gamma^{(0)}/\rho_0$ and $a_1'(\rho_0)\Gamma^{(1)}$ 
($D'=1$ and $a_0(\rho_0)=0$ for this model). 

For the sake of further illustrating the validity 
of Eq. (\ref{eq47}) I will deduce the surface tension as a function of density 
$\rho_0$ near the complete wetting transition, $\rho_0\approx \rho_w$. 
The adsorption coefficients diverge logarithmically near the transition point,
according to
\begin{eqnarray}
\Gamma^{(k)}\sim \nu_k\ln t +\sigma_k,
\label{eq49}
\end{eqnarray}
where $t=1-\rho_0/\rho_w$. Considering the Zwanzig model for prolate  
parallelepipeds in the Onsager approximation, 
for which all $a_k$ are zero except $a_1=4\rho_0/3$, and 
inserting (\ref{eq49}) into (\ref{eq47}) we have 
\begin{eqnarray}
\frac{d\gamma}{d\rho_0}\sim -(\frac{\nu_0}{\rho_0}+\frac{4}{3}\nu_1)
\ln t-\frac{\sigma_0}{\rho_0}-\frac{4}{3}\sigma_1.
\label{eq50}
\end{eqnarray}
Integrating (\ref{eq50}) with respect to $\rho_0$ we obtain
\begin{eqnarray}
\Delta \gamma=\nu_0{\cal L}(t)
-\sigma_0\ln(1-t)+t\left[\tilde{\nu}_1(\ln t-1)+\tilde{\sigma}_1\right] ,
\label{eq51}
\end{eqnarray}
where $\Delta \gamma=\gamma-\gamma_w$, with $\gamma_w$ the surface 
tension value at the transition point, 
$\tilde{\nu}_1=4\rho_w \nu_1/3$, $\tilde{\sigma}_1=4\rho_w \sigma_1/3$, 
and ${\cal L}(t)=\int_0^tdx(1-x)^{-1}\ln x$.

I will use Eq. (\ref{eq51}) for testing the validity 
of (\ref{eq47}) with the exact surface tension calculations near the 
wetting transition.

\section{Polydisperse systems with finite number of particles}
\label{finite}

In real systems the total number of particles although large is a finite 
number. For one-component systems is completely justify to take the 
thermodynamic limit but this is not the case for polydisperse systems. 
The fundamental reason of this difference is that the finite systems 
have only a few species with characteristic lengths distributed in the tail 
(far from the mean value) of the size
distribution function. The majority of theoretical calculations on  
polydisperse systems have in common the hypothesis that for any length 
there are an infinite number of particles (in the thermodynamic sense) distributed
according to the so called parent size distribution function $p(l)$ (normally 
infinite ranged). Although this 
hypothesis is in general not justified the majority of the obtained results 
reflect qualitatively the experimental behaviour because normally the decaying 
law of $p(l)$ for large $l$'s is so rapidly that as is shown here the particles 
with large lengths don't affect too much the properties of the different 
coexisting phases. 
For low decaying $p(l)$'s like the 
log-normal distribution the species with large lengths have a strong influence 
in the phase behaviour of the system. Then, the phase diagrams obtained 
through theoretical calculations using the preceding hypothesis  
can differ very much from the real ones. For this reason becomes crucial 
to estimate the number of particles present in the tail of $p(l)$ and if is 
highly enough to justify the thermodynamic limit. 

Suppose we have a polydisperse 
infinite system described with infinite ranged size 
distribution function $p(l)$ and that we take a sample of $N$ particles 
(with $N$ very large) which coincide with our real system. 
Then, there is a maximum extreme value $L_{\rm{max}}$ for the lengths of 
particles which obviously depend on $N$ and on the degree of polydispersity.
I will obtain this relation using the asymptotic properties of 
sampling distributions (for more information see the reference \cite{Cramer}).
Arranging the $N$ sample values in order of magnitude and considering the
$N_t$'th value from the top, for $N_t=1$ we obtain the extreme value. If 
$l$ denote the length of the $N_t$'th specie from the top, 
the probability element that 
among the $N$ sample values, $N-N_t$ are less than $l$, and 
$N_t-1$ are greater than $l+dl$, while the remaining value falls 
between $l$ and $l+dl$ is:
\begin{eqnarray}
g_{N_t}(l)dl=NC^{N-1}_{N_t-1}F(l)^{N-N_t}\left(1-F(l)\right)^{N-1}p(l)dl,
\end{eqnarray}
where $F(l)=\int_0^ldtp(t)$ and $C^{n-1}_{\nu-1}$ is the combinatorial number.
Introducing the new variable $\xi=N(1-F(l))$ we have that the 
distribution function that follow this variable in the asymptotic 
limit of large $N$'s is $h_{N_t}(\xi)=\xi^{N_t-1}\exp{(-\xi)}/\Gamma(N_t)$, 
where $\Gamma(x)$ is the Gamma function. 
The variable $l$ depends on $\xi$ and $N$ by the relation 
\begin{eqnarray}
\xi=N\int_l^{\infty}p(t)dt.
\end{eqnarray}
If the asymptotic solution of this equation is $l=f_N(\xi)+{\cal O}(
u(N))$ (where $u(N)$ is the next term in the asymptotic 
expansion) we can 
estimate the mean length (the average is taken over an infinite number of 
samples with fixed $N$) of the $N_t$'th specie from the top as
\begin{eqnarray}
L_N(N_t,\Delta)=\int_0^{\infty}d\xi h_{N_t}(\xi)
f_N(\xi)+{\cal O}\left(u(N)\right),
\end{eqnarray}
where was shown explicitly the dependence of $L_N$ on the degree of 
polydispersity $\Delta$. Fixing the total number $N$, 
the maximun extreme value is 
obtained as $L_{\rm{max}}=L_N(1,\Delta)$. Then, the number of particles 
with lengths in the interval
$[(1-r)L_{\rm{max}},L_{\rm{max}}]$ (with $r<1$)
is the solution of the equation 
\begin{eqnarray}
rL_{\rm{max}}=L_N(N_t,\Delta), 
\label{eq52}
\end{eqnarray}
with respect to $N_t$.
Here I will use two different families of length 
distribution functions. All them have in common that the mean rod 
length is fixed to one ($\int dllp(l)=1$). 
Is imposed also that $\lim_{l\to 0}p(l)=0$, i.e. the probability 
of finding rods with zero length is zero. As a measure of the degree of
polydispersity I select the parameter $\Delta=\sqrt{\sigma^2-1}$, 
where $\sigma$ is the mean
deviation
$\sigma^2=\int dl l^2 p(l)$. The last constraint that I impose to $p(l)$ 
is $\lim_{\Delta\to 0}p(l)=\delta(l-1)$, so is recovered the one-component 
system in the limit of small polydispersities.

The first one is defined as 
\begin{eqnarray}
p(l)&=&\frac{qc_{\nu,q}^{\nu+1}}{\Gamma\left[(\nu+1)q^{-1}\right]}l^{\nu}
\exp{\left[-(c_{\nu,q} l)^q
\right]}, \label{eq53}\\
c_{\nu,q}&=&\frac{\Gamma\left[(\nu+2)q^{-1}\right]}
{\Gamma\left[(\nu+1)q^{-1}\right]}.
\nonumber 
\end{eqnarray}
The number $q\ge 1$ controls the decaying law at large $l$'s, and $\nu\ge0$
controls the degree of polydispersity which maximun value  
\begin{eqnarray}
\Delta_{\rm{max}}=
\sqrt{\frac{\Gamma(q^{-1})\Gamma(3q^{-1})}{\Gamma^2(2q^{-1})}-1}, 
\end{eqnarray}
is reached at $\nu=0$. When $q=1$ (\ref{eq53}) coincide with the Schulz's 
distribution function for which $\Delta_{\rm{max}}=1$. 

The second family is 
\begin{eqnarray}
p(l)=C_{\alpha,q} \exp[-\alpha |\ln (l/l_0)|^q],
\label{eq54}
\end{eqnarray}
where $C_{\alpha,q}$ and $l_0$ depend on $\alpha$ and $q$ and 
are calculated from the conditions $\int dlp(l)=\int dl l p(l)=1$.
Specially for $q=2$ we have the log-normal distribution: 
\begin{eqnarray}
p(l)=\frac{1+\Delta^2}{\sqrt{2\pi\ln(1+\Delta^2)}}\exp{\left\{
-\frac{\ln^2\left[(1+\Delta^2)^{\frac{3}{2}}l\right]}{
2\ln(1+\Delta^2)}\right\}},
\label{eq55}
\end{eqnarray}
where $0<\Delta<\infty$.
Applying the described formalism to the distribution function family 
(\ref{eq53}) we obtain 
\begin{eqnarray}
L_N(N_t,\Delta)\approx \frac{t^{1/q}}{C_{\nu,q}}\left[1-d_0\left(1+
\frac{\mu}{t}+(q-1)d_1\right)\right],
\label{eq56}
\end{eqnarray}
where were neglected terms of order $\sim\ln^{1/q-3}(N)$ and I have defined
$\mu=(\nu+1)/q-1$, $t=\ln[N/\Gamma(\mu+1)]$ and 
\begin{eqnarray}
d_k=\frac{1}{qt}\left(\frac{\gamma_k(N_t)}{2^k}-\mu\ln t\right),
\end{eqnarray}
with $\gamma_k(x)$ the Polygamma functions of order $k$. 

\begin{figure}
\mbox{\includegraphics*[width=2.5in, angle=0]{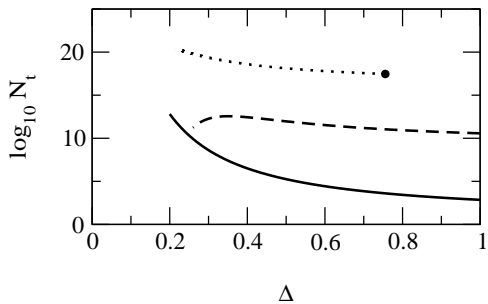}}
\caption[]{Number of species with lengths in the interval 
$[(1-r)L_{\rm{max}},L_{\rm{max}}]$ ($r=4/7$) as a function of 
polydispersity for fixed total number 
of particles: $N=10^{25}$. With solid, dashed 
and dotted lines are shown the results for log-normal, 
Schultz's distribution function 
and $p(l)$ from (\ref{eq53}) with $q=2$ respectively. The point  
represent the value $\Delta_{\rm{max}}$ for $q=2$.} 
\label{fig2}
\end{figure}

For the log-normal $p(l)$ (see Eq. \ref{eq55})  
we obtain 
\begin{eqnarray}
L_N(N_t,\Delta)\approx\frac{\left(1+\Delta^2\right)^{\lambda-1/2}
\Gamma(N_t-1/\lambda)}
{\left(4\pi t\right)^{1/(2\lambda)}\Gamma(N_t)},
\label{eq57}
\end{eqnarray}
where were neglected terms of order of unity.
Now $t=\ln N$ and $\lambda=\sqrt{2t\ln^{-1}(1+\Delta^2)}$.
Fixing $N=10^{25}$ and $r=4/7$ I solved the equation (\ref{eq52}) for $N_t$ 
using $L_N$ from (\ref{eq56}) with $q=1,2$ and from (\ref{eq57}).  
The results ($N_t$ as a function of $\Delta$) 
are shown in figure \ref{fig2}. As we can see from the results 
for low decaying distributions the tails contain less number of particles. 
This behaviour can be understood if we 
take into account that fixing $N$ the probability to find large particles 
strongly decrease with $q$ and then $L_{\rm{max}}$ also decrease in so way 
that the number of particles present in the tail (with length $rL_{\rm{max}}$)  
is an increasing function of $q$.
Another interesting feature is that $N_t$ decrease with $\Delta$ (except 
for $q=1$ and $\Delta<0.3$) which can be explained by the same reason
because $L_{\rm{max}}$ is an increasing function of $\Delta$. The main 
conclusion that we can extract from these results are that the 
thermodynamic limit for the species belonging to the tail of the 
length distribution function is less justified from low decaying 
$p(l)$'s and for large polydispersities.

\begin{figure}
\mbox{\includegraphics*[width=2.5in, angle=0]{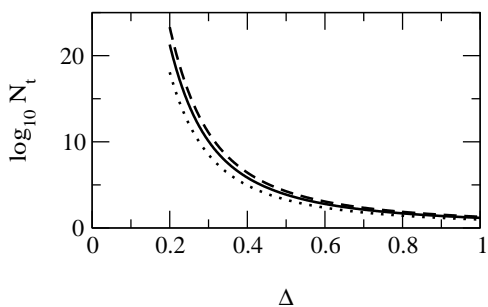}}
\caption[]{Number of species with lengths in the interval 
$[(1-r)L_{\rm{max}},L_{\rm{max}}]$ ($r=4/7$) as a function of 
polydispersity for fixed $L_{\rm{max}}=20,35$ and $50$ (the dotted, solid 
and dashed lines respectively).}
\label{fig3}
\end{figure}

Following the usual theoretical works on polydisperse systems 
I will fix a parent $p(l)$ but taking into account 
that the system contains a finite number of species will be used 
a truncated $p(l)$. The main purpose is to estimate the number of particles 
present in the tails. Following the already described procedure I fix now 
$L_{\rm{max}}$ and solving equation 
$L_N(1,\Delta)=L_{\rm{max}}$ for N I find the dependence $N(\Delta)$.
Obviously each point in the $\rho-\Delta$ phase diagram now represent 
systems with different number of particles. From equation
(\ref{eq52}) I calculate $N_t(\Delta)$ for the same $r=4/7$ 
which is shown in figure 
\ref{fig3} for the log-normal distribution with $L_{\rm{max}}=20,35$ and
$50$. 
Although the relative fraction $N_t/N$ is an increasing function of $\Delta$ 
as should be, 
$N_t$ decrease with polydispersity. Taking into account that the 
shadow phase 
fill an infinitesimal amount of the total volume, the number of its particles 
is less in orders of magnitude compared with 
the cloud phase. Then the fractionation of $N_t\sim 10^3$ 
particles from the parent to the shadow phase 
can change dramatically the mean length.  
We can see from  figure 
\ref{fig3} that until $\Delta\sim 0.6$ we have $N_t\gtrsim 10^3$ so the 
results presented in the next sections for the log-normal distribution function 
should be qualitatively correct in this range of polydispersity values. 

\section{Results}
\label{results}
The important quantities to characterize the surface behaviour of our
hard rod system interacting with hard wall are the 
$k$-moment adsorption coefficients ($k=0,1$)
and the surface tension (see Eqs. (\ref{eq41}), (\ref{eq42}) 
and (\ref{eq43})). 
This quantities are calculated after minimization of (\ref{eq24}) to obtain 
the equilibrium density profiles $\rho_{\mu}(z)$ (see Eqs. (\ref{eq31})-
(\ref{eq33})). 
However, we can gain insight into
the surface phase diagram for this model without carrying out the  
full minimization as I will show bellow.  

Because the hard wall does not allow a rod with length $l$ 
perpendicular to it to be 
at a distance less than $l/2$, we have $m_z(0)=0$, or equivalently 
$\vartheta(0)=
-2\rho_0/3$. Then Eqs. 
(\ref{eq28}) and (\ref{eq29}) 
can be solved to obtain  $\tau(0)$ and $s(0)$, which yield  
inserted in (\ref{eq31}) and (\ref{eq32}) 
$\rho_x(0)$ and $\rho_y(0)$. 
Finally,  $\rho_z^{c}= \int dl \rho_z(l,l/2)$,  
the total contact density of rods perpendicular to the wall,
is calculated using the contact value theorem 
valid for any mixture of particles interacting via hard core. 
This theorem for a mixture of hard core particles reads
\begin{eqnarray}
\beta \Pi=\sum_{\mu} \rho_{\mu}(l_{\mu}/2,l).
\end{eqnarray}
The extension of this theorem to the polydisperse Zwanzig model in the 
Onsager limit is 
\begin{eqnarray}
\beta \Pi=\sum_{\mu=x,y,z} \int dl \rho_{\mu}(l,l_{\mu})=
\rho_x(0)+\rho_y(0)+\rho_z^c,  
\end{eqnarray}
where $\beta \Pi=\rho_0(1+2\rho_0/3)$ is the bulk pressure 
and $l_x=l_y=0$, $l_z=l/2$.
Finally, it is necessary to know for which value of the 
bulk isotropic density $\rho_0$ there appears a macroscopically thick layer 
of the nematic phase at the wall-isotropic 
interface. For complete wetting this value  
can be calculated from (\ref{eq16}) and (\ref{eq17}) 
with $x=1$ (the fraction of volume occupied by the coexisting 
nematic phase is zero).

Having  all this information about the surface structure we can determine 
the possible scenarios of surface phase transitions occurring 
before the wetting transition, depending on 
the form of the length distribution function $p(l)$. For example, 
we can distinguish the nature of the surface transition (second or 
first order). For the former we obtain 
the exact value of the transition point $\rho_0^{\ast}$ using  
equation (\ref{eq40}), and for the latter we can also calculate 
the exact value if we use  the local model, and produce an estimation 
for the nonlocal model, as we will show later.  

I have found that the 
surface phase behaviour depends on the decaying law at large lengths 
of the length distribution function $p(l)$. I have found 
two qualitatively 
different types of surface phase diagrams, one of them coming 
from  $p(l)$'s decaying  
exponentially or faster, and the other 
from slowly decaying $p(l)$'s. 

\subsection{Bulk and surface phase diagrams for rapidly decaying  
distribution functions} 
\label{rapidly}
\begin{figure}
\mbox{\includegraphics*[width=2.5in,angle=0]{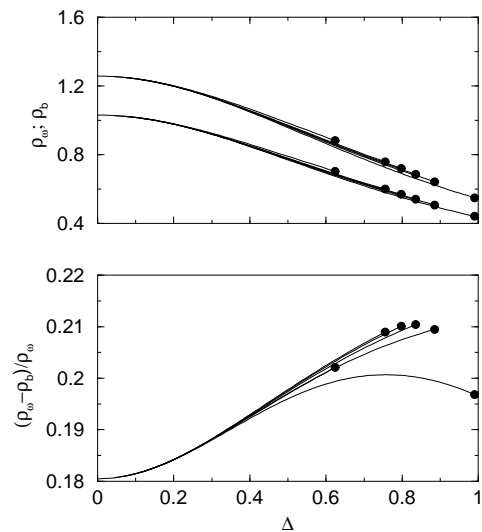}}
\caption[]{In the upper figure the density $\rho_b$ of the  
transition to the biaxial phase and the wetting density 
$\rho_w$ ($\rho_w>\rho_b$) are plotted as a function 
of polydispersity $\Delta$, 
for $q$=1, 1.3, 1.5, 1.7, 2 and 5. 
The bottom figure shows the relative stability of the biaxial phase for the 
same $q$'s.
The points show the maximun degree of 
polydispersity for a given $q$. 
In decreasing order of $\Delta$ 
the points correspond to $p(l)$'s with increasing order of $q$. In all figures  
the densities, first moments and specie lengths  
are dimensionless (see the text 
in section \ref{model}).}
\label{fig4}
\end{figure}

I begin the study with $p(l)$ from the first family (\ref{eq53}). Taking 
into account that I pretend to study systems with very large but finite 
number of particles (see the discussion done in the preceding 
section) the correct way to do this is to select the length 
distribution function equal to $p(l)\Theta(L_{\rm{max}}-l)$, where 
$L_{\rm{max}}$ is the maximun extreme value for the lengths of particles and 
$\Theta(x)$ is the Hevisaide function. 
With the constraints $\int_0^{L_{\rm{max}}} dll^kp(l)=1$ ($k=0,1$) the 
equation (\ref{eq53}) keeps the same but substituting  
$\Gamma(x)$ by the Incomplete Gamma function 
$\Gamma[(c_{\nu,q}L_{\rm{max}})^q,x]$ where $c_{\nu,q}$ is now the solution 
of the equation
\begin{eqnarray}
c_{\nu,q}=\frac{\Gamma\left[(c_{\nu,q}L_{\rm{max}})^q,(\nu+2)q^{-1}\right]}
{\Gamma\left[(c_{\nu,q}L_{\rm{max}})^q,(\nu+1)q^{-1}\right]}.
\end{eqnarray}
Using Eqs. (\ref{eq39}) and (\ref{eq40}) I have 
calculated the second 
order uniaxial-biaxial nematic transition 
(or isotropic-nematic transition in the plane of the wall),
i.e. the bulk density value $\rho_b$ at which  the 
solution $m_x(0)\neq m_y(0)$ bifurcates, 
using this family of $p(l)$'s for several values of $q$.
When the bulk density $\rho_0$ is greater than this value we find a phase with 
preferential alignment of rods parallel to the $x$ axis, i.e. 
$\rho_x>\rho_y,\rho_z$ with $\rho_y\neq \rho_z$ due to the presence of 
the wall.  
This is the reason why in \cite{Roij1} this phase 
was called biaxial nematic phase.
All results using the truncated length distribution 
function from the family (\ref{eq53}) and $q\ge 1$ 
with $L_{\rm{max}}>20$ are 
indistinguishable from the infinite ranged $p(l)$ which shows that for 
rapidly decaying $p(l)$'s the rods with lengths distributed in the 
tail of $p(l)$ don't affect the phase behaviour of the polydisperse 
mixture. 
In Fig. \ref{fig4} the values $\rho_b$ 
for which this transition occur are plotted as a function of 
the polydispersity $\Delta$,  
for different choices of $q$ using the infinite ranged $p(l)$. 
Also are plotted the cloud isotropic curves (the values of $\rho^{(I)}$ 
when $x=1$) as the solution of the 
coexistence equations (\ref{eq16}) and (\ref{eq17}). 
For the complete wetting
transition at the wall-isotropic interface  
the cloud isotropic density $\rho^{(I)}$  
coincide with the wetting density value $\rho_w$. 
As we can see from the figure, 
both curves are practically the same for any distribution 
function $p(l)$ belonging to the family (\ref{eq53}) with $q\ge 1$.
The general trend is the decrease of the transition densities with 
polydispersity. This is due to the fact that  
the alignment of long rods parallel to the wall is entropically favored. 
The amount of long rods increases upon increasing polydispersity, therefor  
this entropic effect increases the total first moment parallel 
to the wall $m(0)=m_x(0)+m_y(0)$ reaching the threshold value for which 
a second order nematic transition appears at the wall. 

Figure \ref{fig4} also shows the relative stability of the biaxial 
nematic phase with respect to the wetting transition, measured by the magnitude 
$1-\rho_b/\rho_w$. As we can see, the stability does not change too much 
(from 0.18 to 0.21) and in general is an increasing function of 
$\Delta$. This means that the bulk transition density $\rho^{(I)}=\rho_w$, 
is more sensitive to the change in polydispersity than the surface 
biaxial transition density $\rho_b$. 

\subsection{Bulk and surface phase diagram for slowly decaying 
distribution functions.}
\label{slowly}
The other selected family of length distribution functions 
$p(l)\Theta(L_{\rm{max}}-l)$ with $p(l)$ from (\ref{eq54}) and with  
$C_{\alpha,q}$ and $l_0$ 
calculated from the usual constraints (the  
equality to one of the zero and first moment). 
There are experimental works with oil-in-water 
emulsions where the droplet size distribution is well described 
by a log-normal distribution function \cite{McDonald}. 
Also was proposed a simple 
statistical model to describe the side 
distribution of nuclei in films  
which grow by the absorption of atoms 
which arrive either by impingement or via surface diffusion 
(example 
is adsorption of Ag on amorphous carbon \cite{Faure})  
The resulting length 
distribution function is log-normal \cite{Granqvist}. In some polydisperse 
homopolymer systems in a single solvent was found untypical topology 
of the cloud-point curves \cite{Konigsveld}. Was concluded on the basis 
of experimental evidence that this behaviour is due to high asymmetry in 
the molecular weight distribution, conclusion that agree with theoretical 
calculations using the Flory-Huggins theory for polydisperse polymers 
with log-normal distribution function \cite{Solc}. As we can see this 
type of distribution is not rare in polydisperse systems with particles 
which dimensions are determined by some coalescence process.

In the set of equations that 
determine the coexistence between the isotropic and nematic polydisperse 
phases there are involved integrals of the type $\int dl l^k p(l) 
\exp(\beta l)$, with $k=0,1$. To calculate the first moment or the density 
of the component parallel to the director in the nematic 
phase we find integrals of this type with  $\beta >0$. 
Then for $p(l)$ decaying at large distances 
slower than an exponential these integrals diverge when 
$L_{\rm{max}} \to \infty$. 
This means that in the phase diagram the unique stable 
phase is the nematic at any density.
When $L_{\rm{max}}$ has a finite value there exist a surface 
phase diagram very different 
for the preceding one, as I will show bellow.

The bulk phase diagram of the polydisperse hard rod system 
with the length distribution function family (\ref{eq54})  
turns out to be more complex, so I will now 
describe in more details the whole bulk phase diagram and not only 
the cloud isotropic branch (or wetting transition line), as in the preceding 
case.

\begin{figure}
\mbox{\includegraphics*[width=2.5in, angle=0]{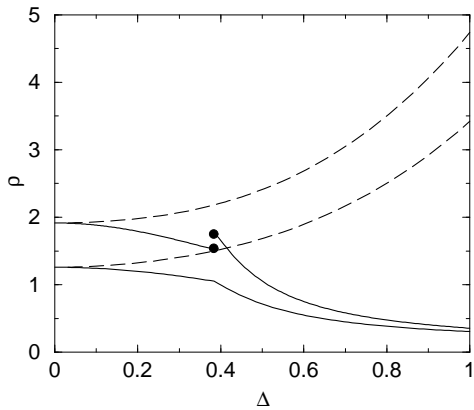}}
\caption[]{Bulk phase diagram of the log-normal polydisperse hard rod 
model with $L_{\rm{max}}=35$ (see Eq. \ref{eq54} with $q=2$). 
The solid lines are the densities of the cloud isotropic ($\rho^{(I)}$) 
and shadow nematic ($\rho^{(N)}$) coexisting phases as a function of 
polydispersity ($\Delta$). The cloud nematic and shadow isotropic are 
the dashed lines. The former exhibit a triple coexistence at 
$\Delta=0.3834$ between an isotropic phase and two different nematic
phases whose densities are marked by black circles in the shadow 
nematic curve.}
\label{fig5}
\end{figure}

Carrying out the same calculations for coexistence as in the previous 
case, but now using (\ref{eq54}) 
with $q=2$ and $L_{\rm{max}}=35$, (i.e. the truncated 
log-normal distribution function), 
we find the bulk phase diagram plotted in Fig. \ref{fig5}.
In this figure the cloud isotropic-shadow nematic ($x=1$) 
and the cloud nematic-shadow isotropic ($x=0$) densities are shown
as a function of $\Delta$.

\begin{figure}
\mbox{\includegraphics*[width=2.5in, angle=0]{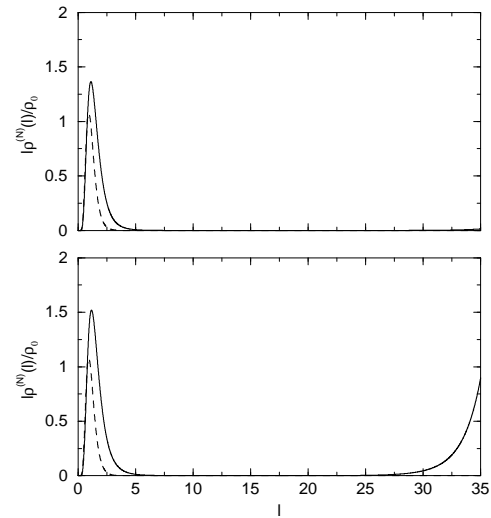}}
\caption[]{The first moment distribution function $l\rho^{(N)}(l)/\rho_0$ 
for the two coexisting nematics. 
Continuous lines represent   
the $N_1$ (top) and $N_2$ (bottom) distribution functions while the  
isotropic cloud distribution is plotted with a dashed line.}
\label{fig6}
\end{figure}

As we can see from the figure there is a polydispersity value 
($\Delta=0.3834$) for which in the cloud isotropic-shadow nematic 
curves appear
a triple coexistence between an isotropic an two different nematic phases 
(a triple point). This behaviour is peculiar of distribution functions 
decaying slower than an exponential for large $l$'s and strongly depends on 
the cutoff $L_{\rm{max}}$. 
I have found that the two coexisting nematics have 
different distribution functions. The nematic with the highest 
density and first moment ($N_2$) has a bimodal first moment 
distribution function $l \rho^{(N)}(l)/\rho_0$ 
with the two values of its maxima having the same order of magnitude, 
whereas the other nematic 
($N_1$) has a unimodal  distribution (see Fig. \ref{fig6}). 

\begin{figure}
\mbox{\includegraphics*[width=2.5in, angle=0]{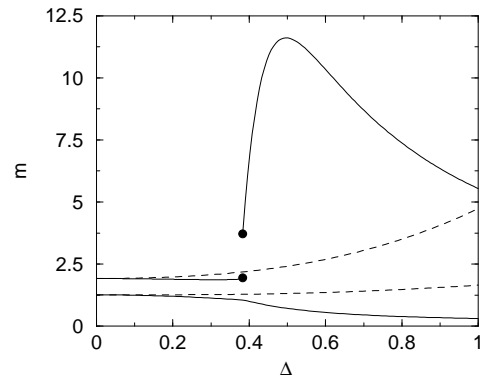}}
\caption[]{The same as Fig. \ref{fig5}, but instead of $\rho$  
the total first moment $m$ is plotted as a function of polydispersity. 
The meaning 
of the different lines is the same as in Fig. \ref{fig5}}
\label{fig7}
\end{figure}

In Fig. \ref{fig7} I have plotted the total moment $m$ of the two 
coexisting isotropic and nematic phases as a function of polydispersity. 
The most interesting feature is the prominent maximun exhibited by the shadow 
nematic curve. Its existence will be explained 
below.

The origin of the second maximum in the function $l\rho(l)/\rho_0$ 
at large polydispersities is very 
easy to explain if we take into account that from Eq. (\ref{eq21}), 
the total moment of the coexisting nematic phase for $x=1$ is proportional 
to the integral
of a term which, for large $l$'s, can be approximated by 

\begin{eqnarray}
f(l)=l\exp{\left(-\alpha\ln^2(\frac{l}{l_0})+4\Delta m l\right)},
\label{eq58}
\end{eqnarray}
where $\Delta m=m^{(I)}_{\perp}-m^{(N)}_{\perp}>0$.
The minimum value of this expression  
is reached at $l^{\ast}$, the largest solution to $f'(l^{\ast})=0$, i.e. 
\begin{eqnarray}
l^{\ast}=\frac{2\alpha\ln(l^{\ast}/l_0)-1}{4\Delta m}.
\label{eq59}
\end{eqnarray}
If $l^{\ast}<L_{\rm{max}}$ we find that the function 
$f(l)$ 
has a second maximum at $l=L_{\rm{max}}$, apart the other one near $l_0$. From 
Eq. (\ref{eq59}) we conclude that increasing polydispersity (decreasing 
$\alpha$) $l^{\ast}$ decreases, so the value $f(L_{\rm{max}})$ increases as 
we can see directly from (\ref{eq58}).
Also from (\ref{eq58}) and the condition 
$l^{\ast}<L_{\rm{max}}$ it is clear that
$f(L_{\rm{max}})$ is an increasing function of $\Delta m$, thus taking into 
account that $m_{\perp}^{(N_2)}< m_{\perp}^{(N_1)}$ 
and $m_{\perp}^{(I_1)}=m_{\perp}^{(I_2)}$, i.e. the most ordered nematic has 
less amount of rods perpendicular to the nematic director, we find that 
for $N_2$  the second maxima value is higher. 
For the less ordered 
nematic ($N_1$), although $l\rho^{(N)}(l)/\rho_0$ 
can develop a second maximum 
at $L_{\rm{max}}$, its value is less than the first maximum by several orders 
of magnitude. This is the reason why we call the $N_1$ distribution 
unimodal distribution.

The maximun in the shadow nematic curve (total moment $m$ 
as a function of $\Delta$ (see Fig. \ref{fig7})) 
as function of $\Delta$ can be explained again by resorting to Eq. (\ref{eq58}). From  
the general cloud isotropic curve behaviour as a function of polydispersity 
$\Delta m(>0)$ is a decreasing 
function of $\Delta$.  
From (\ref{eq58}) the value of $f(l)$ for large $l$'s depends on 
the two terms in the exponential. For moderate polydispersities (near 
the triple point) a rapid decrease in $\alpha$ determines the increase 
of $\int dl f(l)$, whereas for large polydispersities, where $\alpha$ 
changes slowly with $\Delta$ and $\Delta m$ keeps on decreasing, the 
second term dominates the behaviour of this integral, which decreases with 
$\Delta$. Thus at intermediate values $m$ pass trough a maximun.

When $\Delta$ is larger than its triple point value 
the stable nematic is the bimodal one. The cloud nematic and 
the shadow isotropic curves ($x=0$) 
have the usual behaviour with polydispersity, in 
the sense that they show, for any $\Delta$, a single I-N coexistence,
so is expected that for a certain $x$ between 1 and 0 
this triple coexistence will disappear. 

\begin{figure}
\mbox{\includegraphics*[width=2.5in, angle=0]{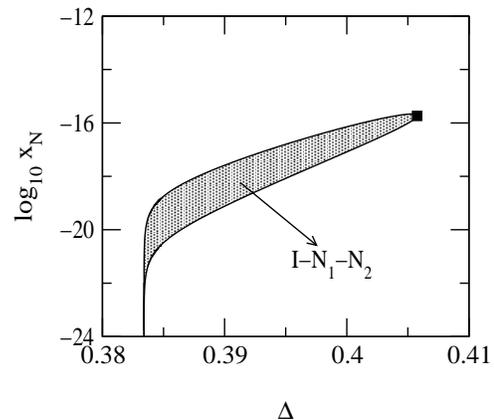}}
\caption[]{The limited area by the closed curve 
in the $x_N-\Delta$ plane coincide with the triple coexistence region which 
finish at a tricritical point (the filled square).} 
\label{fig8}
\end{figure}

From equation (\ref{eq15}) we can estimate the value $x$ for which the 
contributions of the two coexisting phases to the integral have 
the same order at $l=L_{\rm{max}}$ (the length value for the second maxima 
in the $N_2$ first moment distribution function). I obtain 
$x\approx 1-3\exp{(-4\Delta m L_{\rm{max}}/3)}$.
Taking into account that $L_{\rm{max}}=35$ and that $\Delta m$ 
is of order of unity we conclude that the range of $x_N\equiv 1-x$ 
for which take place the preceding described behaviour 
should be of order $10^{-20}$. Thus the triple coexistence 
will extend until very small values of $x_N$. 
To find the triple equilibrium I have solved the set of five equations: 
$s_\alpha=\rho_0\phi_{s_{\alpha}}$, 
$\tau_{\alpha}=\rho_0\phi_{\tau_{\alpha}}$ (with $\alpha=N_1,N_2$) 
and $\Delta \Pi(\tau_{N_2},s_{N_2})=0$ 
(see equations (\ref{eq16}), (\ref{eq17}) and 
(\ref{eq22})) whereas $\rho_0$ is calculated from $\Delta \Pi
(\tau_{N_1},s_{N_1})=0$, i.e. from equation (\ref{eq23}) with 
substitution $N\to N_1$.  The unknowns are $s_{\alpha}$, $\tau_{\alpha}$ and
$x_N$. The definition of $E(l)$ is now 
\begin{eqnarray}
E(l)=3(1-x_N)+x_{N_1}\epsilon_1(l)+x_{N_2}\epsilon_2(l),
\end{eqnarray}
where $x_N=x_{N_1}+x_{N_2}$ ($x_{N_{\alpha}}=x_N\theta_{\alpha}$, with 
$\theta_1+\theta_2=1$) is the total fraction of volume 
occupied by the nematic phases and $\theta_{i}$ measure the relative 
fraction of the $N_i$ phase with respect to the other.  
In figure \ref{fig8} is plotted in the $x_N-\Delta$ plane the closed 
area where the triple coexistence was found. Fixing some $\Delta$ 
in the range $[0.3834,0.4058]$ and increasing $x_N$ from zero is 
reached the value for which appear $I-N_1-N_2$ triple coexistence 
with infinitesimal amount of the $N_1$ phase ($\theta_1=0$). 
Changing the relative 
fraction of the nematic $N_1$ with respect to the $N_2$ we move 
through the triple 
coexistence line which finish when the nematic $N_2$ fill an infinitesimal 
amount of volume ($\theta_1=1$). 

\begin{figure}
\mbox{\includegraphics*[width=2.5in, angle=0]{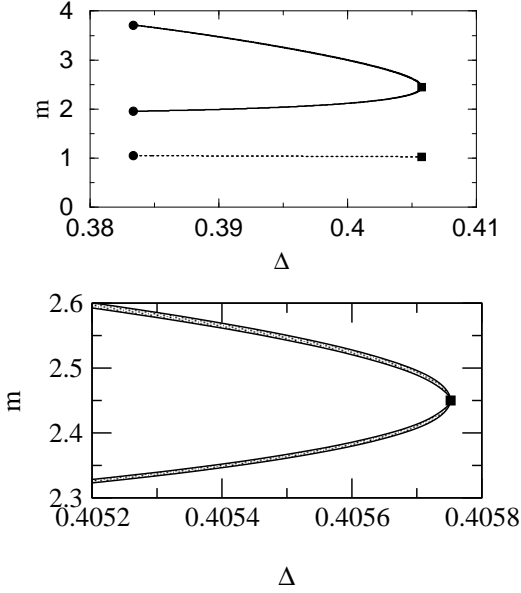}}
\mbox{\includegraphics*[width=2.7in, angle=0]{fig10.eps}}
\caption[]{Phase diagram which corresponds 
to the triple coexistence region of figure \ref{fig8}. $m^{(N_{\alpha})}$ 
as a function of $\Delta$ (the solid line)
are the different branches which begin 
at $\Delta=0.3834$ (the filled circles) and join at the 
tricritical point (the filled square). The dashed line represent 
the values for $m^{(I)}$.   
The bottom figure is a 
zoom of the top one around the tricritical point.}
\label{fig9-10}
\end{figure}

The first moment values for the three phases with $x_N$ and $\Delta$ 
inside the triple 
coexisting region are plotted as function of $\Delta$ in figure 
\ref{fig9-10} where the bottom figure is a zoom of the top figure 
for values of $m$ corresponding to the coexisting 
nematics near the tricritical point.  
The shaded area has the same meaning as in 
figure \ref{fig8}, i.e. represent the allowed values of $m$ inside 
the triple coexisting region. As we can see the gap between the values of $m$
corresponding to coexisting  nematics   
decrease monotonically with $\Delta$
from the triple point corresponding to $x_N=0$ (the filled circles) 
until the tricritical
point for $x_N\sim 10^{-16}$ (the filled square).  

\begin{figure}
\mbox{\includegraphics*[width=3.0in, angle=0]{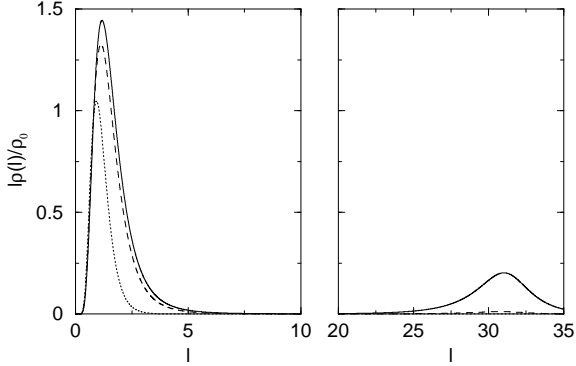}}
\caption[]{The first moment nematic  
distribution functions $l\rho(l)/\rho_0$ for $\Delta=0.395$ 
corresponding to the triple coexistence with 
$x_{N_2}=x_N$.
$N_2$: solid line, $N_1$: dashed line and  $I$: dotted line.}
\label{fig11}
\end{figure}

In figure \ref{fig11} are plotted the first moment distribution 
function $l\rho(l)/\rho_0$ of the three coexisting
phases for $\Delta=0.395$ and $x_{N_2}=x_N$. As we can see the phase 
$N_2$ has the second maxima at $l< L_{\rm{max}}$.  

\begin{figure}
\mbox{\includegraphics*[width=2.5in, angle=0]{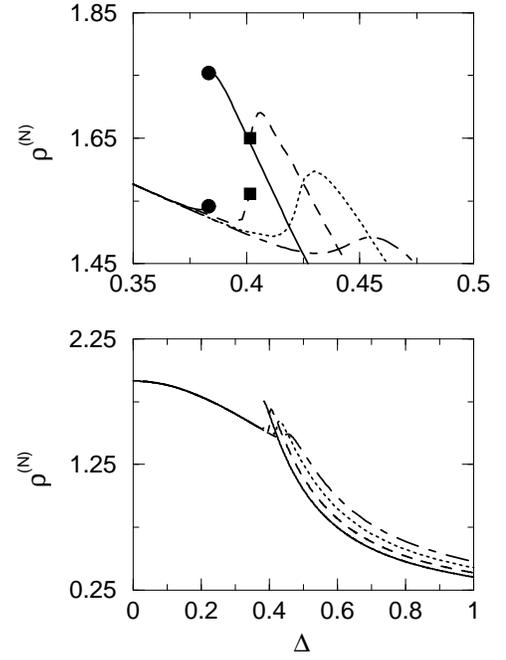}}
\caption[]{The shadow nematic curves for different $L_{\rm{max}}$'s. 
In the top: an interval of $\Delta$ near the triple point. In the  
bottom: the whole range of $\Delta$. 
In both figures $L_{\rm{max}}=35$: solid line, 
$L_{\rm{max}}=30$: dashed line, $L_{\rm{max}}=25$: dotted line 
and $L_{\rm{max}}=20$: dot-dashed line. The circles and squares represent the 
coexisting nematic phases at the triple points.}
\label{fig12}
\end{figure}

I also studied the effect that has the change of the maximun extreme value 
$L_{\rm{max}}$ on the bulk phase diagram.  
In figure \ref{fig12} I plot the nematic shadow curve 
for $L_{\rm{max}}=35$, $30$, $25$ and $20$. As we can see the position 
of the triple point moves to larger $\Delta$'s when $L_{\rm{max}}$ 
decrease, and also 
there is some $L_{\rm{max}}$ (between 30 and 25) 
for which disappear the triple 
coexistence. Is reasonable to predict this behaviour  because the second 
maxima value of the shadow nematic ($N_2$) first moment distribution function 
is an increasing function of $L_{\rm{max}}$ and $\Delta$. 
Then if $L_{\rm{max}}$ decrease 
the value $\Delta$ needed for keeping the triple 
coexistence should increase. We 
conclude also from figure \ref{fig12} (the bottom figure)
that the coexisting densities for $\Delta$'s greater than the triple 
point value decrease with polydispersity when $L_{\rm{max}}$ increase.

\begin{figure}
\mbox{\includegraphics*[width=2.5in, angle=0]{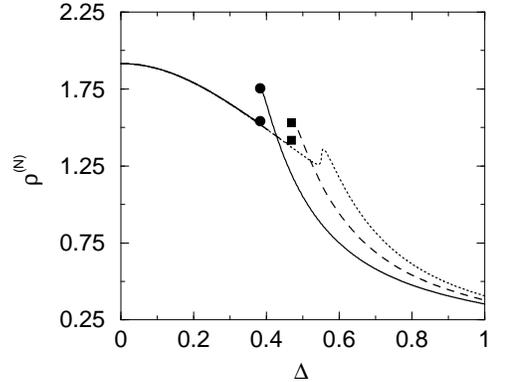}}
\caption[]{The shadow nematic curves for length distribution functions with 
different decaying laws and $L_{\rm{max}}=35$. $q=2$: solid line, $q=2.25$: dashed line and 
$q=2.5$: dotted line.}
\label{fig13}
\end{figure}

The effect of the decaying law of the length distribution 
function for fixed $L_{\rm{max}}$ 
on the bulk 
phase diagram is similar 
to the preceding behaviour as we can see from figure \ref{fig13}.
In  this case was changed the parameter $q$ in the family (\ref{eq54}). 
Increasing $q$ the triple point position moves to larger $\Delta$'s and 
disappear at some value between 2.25 and 2.5. The second 
maxima is higher for low decaying length distribution functions and then 
is necessary less amount of polydispersity to exhibit the $N_1-N_2$ 
transition.

\begin{figure}
\mbox{\includegraphics*[width=2.5in, angle=0]{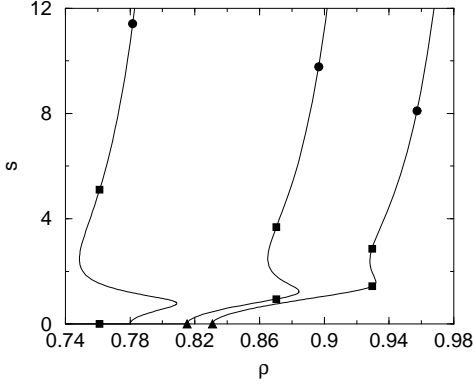}}
\caption[]{$s=m_x(0)-m_y(0)$ as a function of $\rho_0$ for 3 
different values of polydispersity: $\Delta$=0.4705, 0.4298 and
0.4112 from the left to the right}
\label{fig14}
\end{figure}

Now I describe the surface phase diagram for the log-normal 
distribution function.
Fixing $L_{\rm{max}}=35$ the values for the surface parameter $s=m_x(0)-m_y(0)$ 
as a function of $\rho_0$ (the 
bulk density at infinite distance from the hard wall) obtained 
as the solution of the equations (\ref{eq28}) and (\ref{eq29}) 
for different values of 
$\Delta$ are plotted in figure \ref{fig14}.
For small values of polydispersity (less than those shown in the figure) 
the surface phase behaviour is 
similar to the characteristic behaviour of the fast decaying distribution  
functions  where at some 
$\rho_0$ (in figure this value is determined by the intersection 
of the curve with the $\rho$ axis) 
appears a biaxial nematic phase through a second order transition. Then 
the order parameter $s$ increase with density until reaching the  
complete wetting density value (these transition points are marked with 
black circles). 

When $\Delta$ is larger than some value $\Delta^{\ast}\approx 0.4$ 
the graphics of $s$ as 
a function of $\rho$ develop an hysteresis branch. 
This is an indication of 
a first order transition between two nematic phases at the wall. This 
behaviour is common for all truncated length distribution function with slow 
decay at large $l$'s and have not analog in the distribution function family 
with fast decay as we have already seen.
In figure \ref{fig14} are shown three of these hysteresis branches for different
$\Delta$'s. The scenarios is the following: when 
$\Delta>\Delta^{\ast}$ after the second order 
transition from isotropic in plane phase ($m_x(0)=m_y(0)$) 
to the nematic in plane phase 
(the triangles at  figure 
\ref{fig14} represent the bulk density value for which this 
transitions occur) at some point occur a first order surface  
transition at the wall 
between two nematics (see the squares in figure 
\ref{fig14}) and then for larger $\rho$ 
the complete wetting  transition (the circles 
in figure \ref{fig14}). The curves in figure \ref{fig14} 
for $\Delta=0.4112$ and $0.4298$ represent 
this behaviour. Increasing $\Delta$ the first order 
nematic-nematic transition density value move to the second order 
isotropic-nematic transition value and then
for some $\Delta^{c}$ these two transition points coalesce. 
For $\Delta>\Delta^{c}$ the surface first order transition occur 
between an isotropic and nematic phase and then the complete wetting  
transition as usual (see in figure \ref{fig14} the curve for 
$\Delta=0.4705$).

The first order transition points can be calculated exactly using 
the local approximation without carrying out the full minimization 
of the grand potential with respect to the density profiles. I describe 
now the followed procedure to find them.
The excess surface free energy density for the non local model 
which integral over ``z'' is the surface 
tension ($\gamma=\int dz \psi(z)$) can be expressed as a
function of $\Delta \rho(z)$ ($\Delta \rho(z)
=\rho_0-\rho(z)$), $\theta(z)$, $\tau(z)$ and $s(z)$ as 
\begin{eqnarray}
\psi=\Delta \rho+
\frac{s^2-\tau^2}{2}-\frac{2}{3}\rho_0(2\tau+\theta)-
\theta\tau.
\end{eqnarray}
The gradient of $\psi(z)$ is 
\begin{eqnarray}
\nabla \psi&=&-\nabla \tau(\theta+\frac{2}{3}\rho_0)-
\frac{2}{3}\rho_0p(2z)e^{-2\int_0^{2z}dz'\tau(z')} \nonumber \\
&-&\frac{2}{3}\rho_0\int_0^{2z} dl p(l)\tau\ast\delta_l^-
e^{-2\tau\ast\theta_l},
\label{eq60}
\end{eqnarray}
where I have introduced the notation
\begin{eqnarray}
\tau\ast\delta_l^-=\tau(z-\frac{l}{2})-\tau(z+\frac{l}{2}).
\end{eqnarray}
The gradient of $\tau(z)$ is related to $\nabla \theta(z)$ by 
\begin{eqnarray}
\nabla \tau=-\nabla \theta \langle l^2\rangle_c\frac{
1-\langle l^2\rangle_c(1-r)}{1-\langle l^2\rangle_c^2(1-r^2)}; \quad 
r=\frac{\langle l^2\rangle_s}{\langle l^2\rangle_c}, \\
\langle l^2\rangle_{s,c}=\frac{2}{3}\rho_0\int dl l^2 p(l)e^{-(\tau+\theta)l}
\left(\begin{matrix}\sinh(ls) \\ \cosh(ls)\end{matrix}\right), \nonumber 
\end{eqnarray}
where
\begin{eqnarray}
\nabla \theta&=&\frac{2}{3}\rho_0
\int_0^{\infty} dl p(l)e^{-2\int_z^{z+l}dz'\tau(z')}\nonumber \\
&-&\frac{2}{3}\rho_0\int_0^zdl p(l)
e^{-2\int_{z-l}^zdz'\tau(z')}.
\label{eq60a}
\end{eqnarray}
For the local model (see eqs. (\ref{eq35}) and (\ref{eq36})) 
the equation (\ref{eq60}) 
transforms to 
\begin{eqnarray}
\nabla\psi(z)=-\frac{2}{3}\rho_0p(2z)\exp{\left(-4\tau(z)z\right)}.
\label{eq61}
\end{eqnarray}
From (\ref{eq61}) we conclude that the excess energy density profile 
$\psi(z)$ for the local model is a monotonic decreasing function of z and that
$\nabla \psi(0)=0$.
In the range of $\rho_0$ values where there are two solutions  
(I will call them 1 and 2) of the set 
(\ref{eq28})-(\ref{eq29}) 
(two nematics or one isotropic and the other nematic) 
the more ordered phase has a higher value of $\tau(z)$ near 
the wall. Then, from (\ref{eq61}) we see that the excess 
energy density profile of the more ordered 
branch has a smaller slope. Then the transition point (the value of $\rho_0$)
can be  calculated exactly
from $\psi_1(0)=\psi_2(0)$ which taking into account the relation 
$\psi(z)=\Pi_0-\Pi(z)=0$ (where $\Pi(z)$ is the local pressure, i.e. the 
pressure evaluated at $\rho_{\mu}(z)$)) becomes in $\Pi_1(0)=\Pi_2(0)$.
The total integral of $\psi(z)$ 
should be minimized to obtain the equilibrium profile, then the length of 
the more ordered phase for higher $\rho_0$'s is calculated from the 
intersection between two branches $\psi_1(z)$ and $\psi_2(z)$. The length of 
this layer increase continuously from $l^*$  as a function of $\rho_0$. 
The value $l^*$ depends on the form of $p(l)$ for small $l$'s. For 
length distribution functions that have a continuous decay to zero 
when $l\to 0$,  
$l^*=0$ and the transition from less ordered to more ordered density 
profile near the wall is continuous, 
but if $p(l)$ is zero for $l<l_0$, then $l^*=l_0$ and 
at the transition point coexist two profiles with different 
adsorption coefficients, one of them including a layer of length $l^*$ of 
more ordered phase, so the transition becomes first order.

\begin{figure}
\mbox{\includegraphics*[width=2.5in, angle=0]{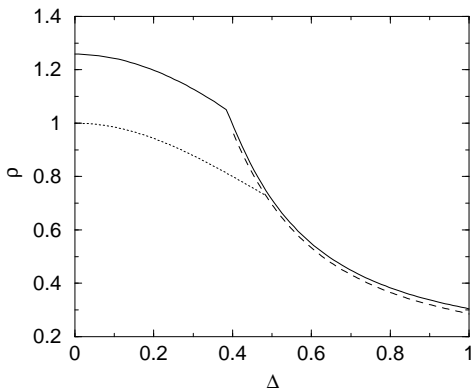}}
\caption[]{The surface phase diagram of the log-normal polydisperse 
hard rod fluid with $L_{\rm{max}}=35$. Solid line: The complete wetting line. 
Dashed line: The first order I-N or N-N at the wall. Dotted line: 
The continuous I-N phase transition.}
\label{fig15}
\end{figure}

\begin{figure}
\mbox{\includegraphics*[width=2.5in, angle=0]{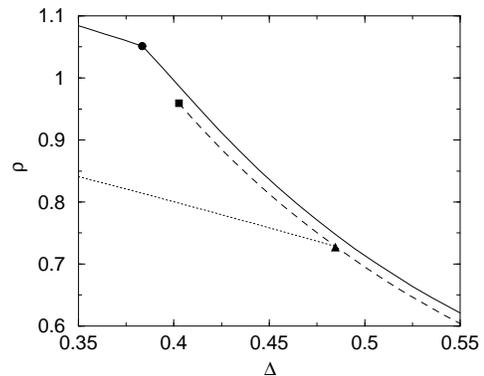}}
\caption[]{The zoom of figure \ref{fig15}. The point is the 
density value of the bulk triple point. The square is the critical 
point of the N-N 2D phase transition. The triangle is the point 
of coalescence of the continuous I-N and a first order I-N}
\label{fig16}
\end{figure}

We conclude that although 
at the transition point there is a ``2D'' first order transition  
between two different phases, in ``3D'' the order of this transition 
depends on the form of $p(l)$ for small $l$'s.
The squares in figure \ref{fig14} were calculated using this procedure 
and the complete surface phase diagram for the log-normal polydisperse 
system with $L_{\rm{max}}=35$ is shown in figures 
\ref{fig15} and \ref{fig16}.
When the polydispersity value is less than the value corresponding 
to the square in figure \ref{fig16} we find after the continuous 
uniaxial-biaxial nematic transition (the transition 
density is shown with dotted line) 
the complete wetting transition (the transition density is 
shown with solid line). 
When $\Delta$ is between the square and the triangle there is also 
a 2D first order N-N transition in between. Finally 
for $\Delta$'s larger than the position of the triangle this 2D first 
order transition is between isotropic and nematic phases. 
Just at the triangle coalesce a second and first order transitions.
The solution of the set of equations (\ref{eq28})-(\ref{eq29}) 
at $z=0$ and the condition 
$\psi_1(0)=\psi_2(0)$ is equivalent 
to find the coexistence between two phases in the two dimensional  
bulk system of polydisperse hard rods ($\rho_z(0)=0$ and $m_z(0)=0$) with 
the excess energy density $\Phi^{exc}=2\Phi_{2D}^{exc}$, where 
$\Phi_{2D}^{exc}$ is the excess part of the bulk free energy density of 
the 2D system.
The reason why I have found two coexisting nematics in the 
bulk phase diagram when $x_N=0$ only for $\Delta=0.3834$ 
(the bulk triple point)
whereas for $\Delta \gtrsim 0.4$ 
exists at the wall a whole range of 
first order N-N transition is the following: 
The equations (\ref{eq28}) and (\ref{eq30})  
for $z=0$ can be put in 
the form
\begin{eqnarray}
\tau=\frac{2}{3}\rho_0\left(\int dl lq(l)e^{-\tau l}\cosh(sl)-1\right), \\
s=\frac{2}{3}\rho_0\int dllq(l)e^{-\tau l}\sinh(sl),
\end{eqnarray}
which form coincide  
exactly with the set of equations resulting from  
the equality of chemical potentials between the cloud isotropic 
and shadow nematic phases for $2D$ system with the following 
variable definitions: $\tau=(m_x^{(2D)}+m_y^{(2D)})/2-m^{(2D)}_0$, 
$s=(m_x^{(2D)}-m_y^{(2D)})/2-m_0^{(2D)}$, $\rho_0^{(2D)}=4\rho_0/3$ and 
$q(l)=p(l)e^{2\rho_0l/3}$. 
Then the surface-parent 
phase from which separate the different coexisting phases have not 
a log-normal distribution form. If the bulk density $\rho_0$ 
is large enough then
the ``surface-parent'' length distribution function $q(l)$  
has a second maxima at $z=L_{\rm{max}}$ and  with a bimodal 
parent distribution function is possible to find a nematic-nematic 
coexistence.  

For the non local model is very difficult to gain insight about the 
monotonic behaviour of the density profiles as can be seen from the equations 
(\ref{eq60})-(\ref{eq60a}). 
In principle they can be non monotonic functions of $z$.
Then, the location of the transition points between different 
surface phases through a first order transition 
may differ from the local model results.  
\subsection{The wall-isotropic and isotropic-nematic interfaces}
\label{w-i-n}
I have solved the equations (\ref{eq28})-(\ref{eq30}) using the Schultz's length  
distribution function for the bulk isotropic phase to obtain the equilibrium 
wall-isotropic profiles. I used the Gauss-Laguerre quadrature to calculate the integrals 
over variable $l$ using 10 points and using the fast Fourier transform 
algorithm to calculate the convolutions $\rho_z\ast \theta_l$ 
for each $l$ with a grid spacing $\Delta z=l_{\rm{min}}/10$ 
($l_{\rm{min}}$ is the smallest quadrature point). 
I have compared the results with those 
obtained by increasing the number of points used in the quadrature 
until 20 and the density profiles 
and the surface tension values does not change appreciably for $0<\Delta
<0.55$.

\begin{figure}
\mbox{\includegraphics*[width=3.5in, angle=-90]{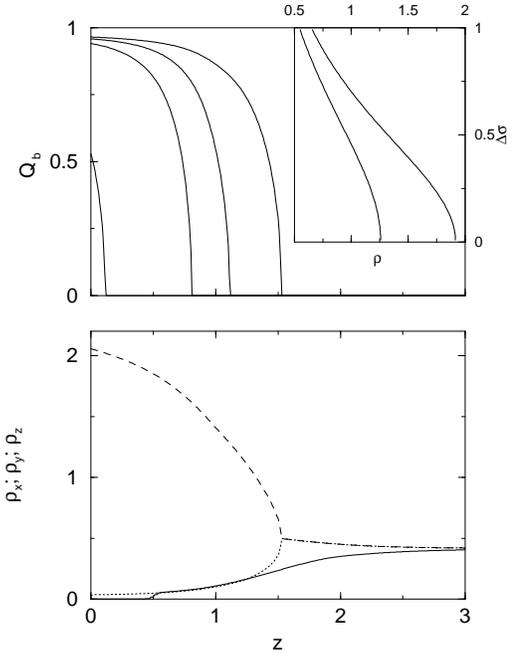}}
\caption[]{The upper figure: Order parameter profile $Q_b(z)$ for 
$\Delta=0.064$ and $\rho_0=1.05$, 1.2, 1.23 and 1.245 (from left 
to right). The inset: 
Biaxial nematic transition density $\rho_b$ and 
the complete wetting density $\rho_w$ as a function of $\Delta$. 
The bottom figure: Density profiles $\rho_x(z),\rho_y(z),\rho_z(z)$ for 
$\rho_0=1.245$.}
\label{fig17}
\end{figure}

\begin{figure}
\mbox{\includegraphics*[width=3.in, angle=0]{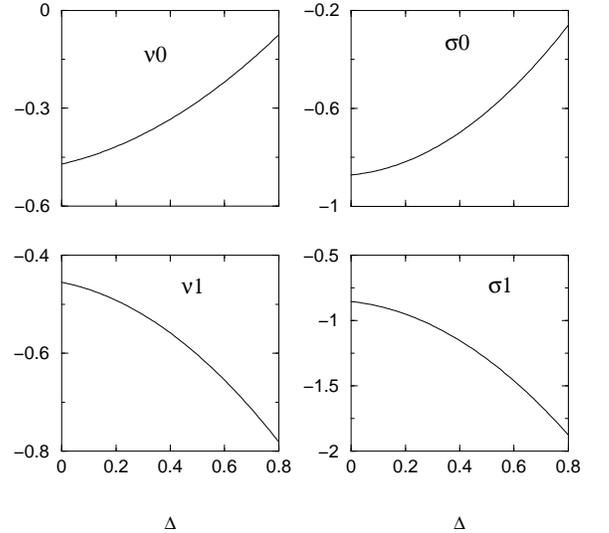}}
\caption[]{The coefficients $\nu_k$ and $\sigma_k$ of the logarithmic divergence 
as a function of polydispersity.}
\label{fig18}
\end{figure}

\begin{figure}
\mbox{\includegraphics*[width=2.5in, angle=0]{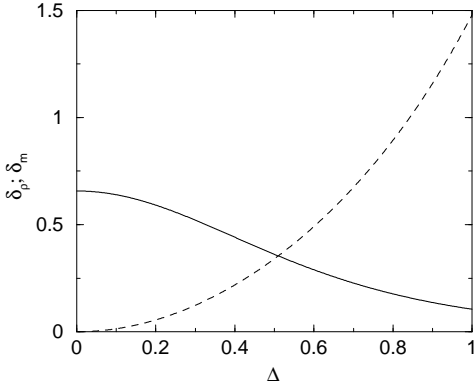}}
\caption[]{The quantities 
$\delta_{\rho}=\rho^{(N)}-\rho^{(I)}$ (solid line) and 
$\delta_m=m^{(N)}-m^{(I)}$ (dashed line) as a function of $\Delta$.}  
\label{fig19}
\end{figure}

\begin{figure}
\mbox{\includegraphics*[width=3.in, angle=0]{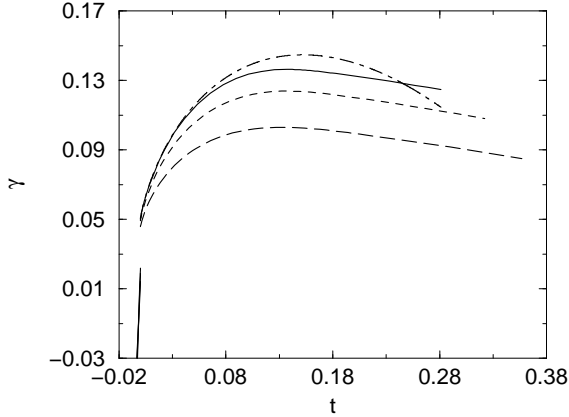}}
\caption[]{The wall-isotropic surface tension as 
a function of $t=1-\rho/\rho_w$ for 
$\Delta=0.064$ (solid line), 0.33 (dashed line) and 0.53 (long line). 
The dot-dashed line represent the analytic approximation 
(\ref{eq51}) for $\Delta=0.064$}
\label{fig20}
\end{figure}

For polydispersity value $\Delta=0.064$ and bulk 
isotropic density $\rho_0=1.02 > \rho_b$ ($\rho_b$ is the density 
of the isotropic-biaxial nematic transition which is 
shown as a function of $\Delta$ in the inset of figure \ref{fig17}) 
I obtain the density profiles $\rho_x(z)$, $\rho_y(z)$ and $\rho_z(z)$ 
shown in figure \ref{fig17}. In the upper figure are plotted the order in 
plane parameter $Q_b=(\rho_x-\rho_y)/(\rho_x+\rho_y)$ for different 
values of $\rho_0$. When $\rho_0 \to \rho_w$ ($\rho_w$ 
as a function of $\Delta$ is also plotted in the inset of the 
upper figure) the point $z^*$ where separates 
the $\rho_x$ and $\rho_y$ branches go to infinity continuously as should 
be in the complete wetting transition, with the final profile 
composed by macroscopic layer of the nematic phase. 
I have checked that   
$\gamma_{WI}=\gamma_{WN}+\gamma_{NI}$ for $\rho_0=\rho_w$
(with $\gamma_{WI}$, $\gamma_{WN}$ and $\gamma_{NI}$ the wall-isotropic, 
wall-nematic and nematic-isotropic surface tensions) which should be 
fulfilled at the complete wetting transition. I also have found that  
$\Gamma^{(0)}$ and $\Gamma^{(1)}$ diverges 
logarithmically with coefficients $\nu_i$ and $\sigma_i$ 
depending on $\Delta$ as we can see 
from figure \ref{fig18}. The general trend is that the prefactor ($\nu_0$)
of the logarithmic divergence of the
density adsorption coefficient is a 
decreasing function (in absolute value) of $\Delta$, whereas the 
prefactor ($\nu_1$) of the logarithmic divergence 
of the first moment adsorption 
coefficient  
is an increasing function (in absolute value) of $\Delta$. This result is due 
to the fact that the density gap between the wetting  
density $\rho_w=\rho^{(I)}$ and the density of the coexisting nematic 
phase $\rho^{(N)}$ 
decrease with polydispersity as we can see from figure \ref{fig19}, whereas  
the moment gap is an increasing function of $\Delta$ (see figure \ref{fig19}). 
The surface tension as a function of $\rho_0$ is plotted in figure 
(\ref{fig20}) for three 
different values of $\Delta$. As I have already shown in 
section \ref{thermo} the behaviour of $\gamma$ is governed by  
certain combination of $\Gamma^{(0)}$ and $\Gamma^{(1)}$ 
(see equation (\ref{eq47})), 
thus taking into account that
these quantities have a logarithmic divergence can be used the 
approximate analytic equation (\ref{eq22}) to estimate the difference 
between the wall-isotropic 
surface tension $\gamma_{WI}$ for $\rho_0<\rho_w$ and its wetting value at 
$\rho_0=\rho_w$. 
For comparison I have plotted in figure 
\ref{fig20} the exact value of $\gamma_{WI}$ and the approximation 
(\ref{eq22}) for $\Delta=0.064$. 
If we take into the account that the comparison between 
them should be restricted to the narrow neighborhood of $\rho_w$ is surprising 
that they are similar until $t=1-\rho/\rho_w \approx 0.1$, and 
that the maximun position is well estimated. Also is plotted $\gamma_{WI}$ 
for $\Delta=0.33$ and $0.53$.

\begin{figure}
\mbox{\includegraphics*[width=2.5in, angle=0]{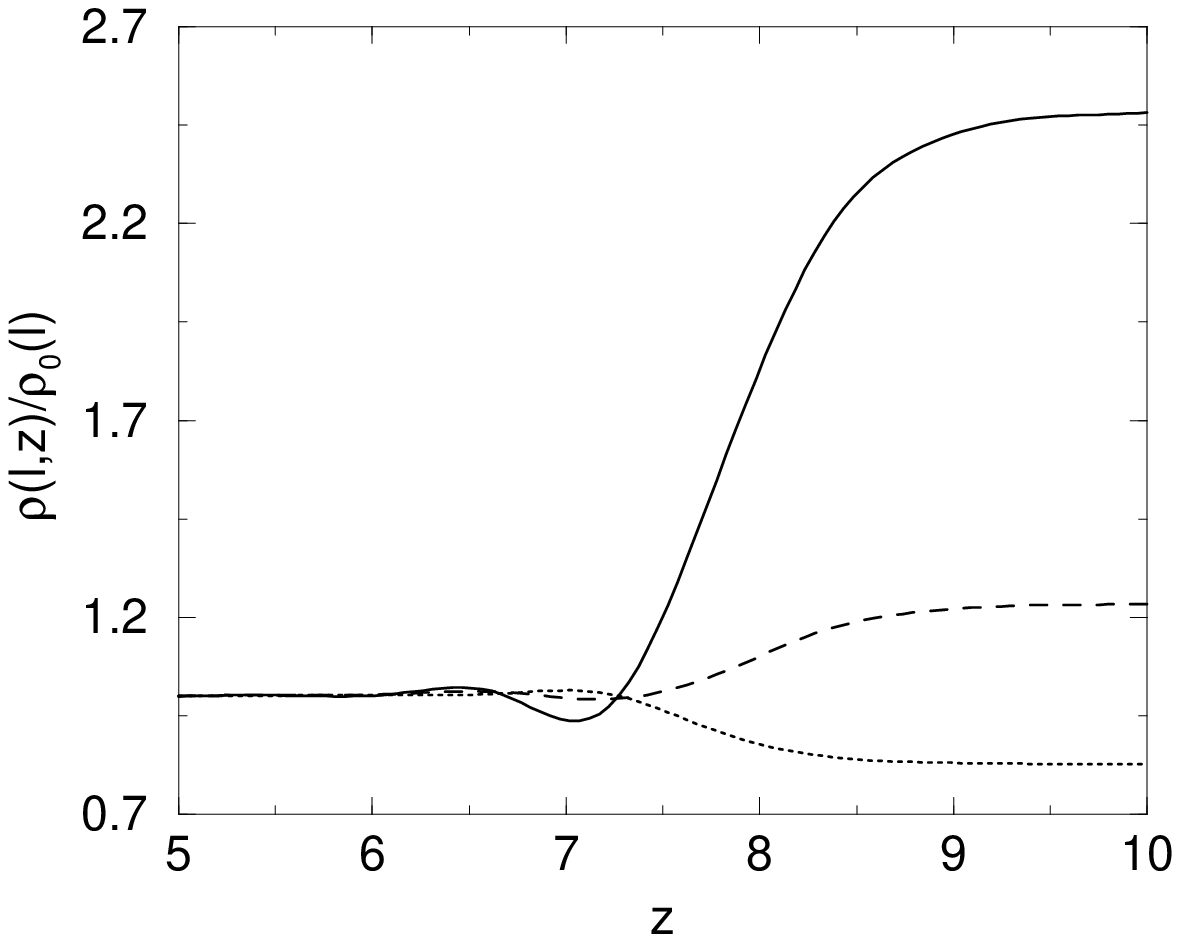}}
\mbox{\includegraphics*[width=2.5in, angle=0]{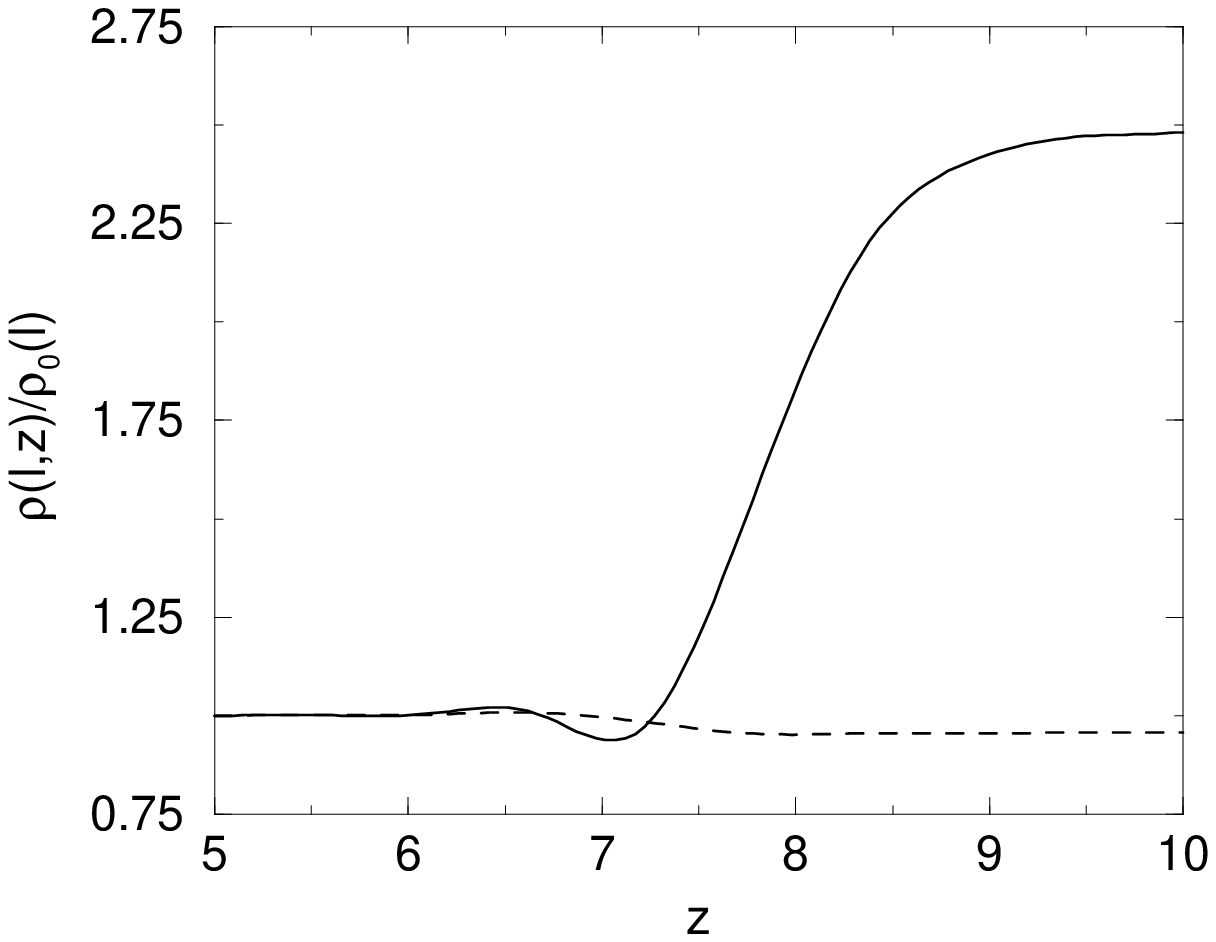}}
\caption[]{The upper figure: the density profile $\rho(l,z)/\rho_0(l)$
of the I-N interface for $\Delta=0.577$. Are plotted three profiles for
different $l$'s. Solid line: $l=1.91$, dashed line: $l=1.21$ and
dotted line: $l=0.67$. The bottom figure: Profiles corresponding
to $l=1.91$ for $\Delta=0.577$ (solid line) and $\Delta=0.333$ (dashed line).}
\label{fig21-fig22}
\end{figure}

In figure \ref{fig21-fig22} (the upper one)
are shown the isotropic-nematic interfaces 
for $\rho_0=\rho_w$ and $\Delta=0.577$.  
In particular are plotted the 
profiles $\rho(l,z)/\rho_0(l)$ for three different $l$'s. 
This ratio should tend to 1 at the bulk isotropic side of the interface. 
From figure is clear that the nematic phase has  
more amount of large rods than the isotropic one.  
This effect 
is more dramatic for higher $\Delta$'s. In the bottom figure I show 
for fixed $l=1.91$ the profiles for two different $\Delta$'s. 
By increasing $\Delta$ 
the amount 
of rods with this length changes its segregation behaviour. For $\Delta=0.333$  
they are lightly segregated at the isotropic phase whereas for $\Delta=0.577$
they strongly segregate to the nematic phase.

I have also calculated the wall-isotropic interface in the local 
approximation using the log-normal  
length distribution function for the bulk isotropic phase solving 
the equations (\ref{eq28}), (\ref{eq29}) and (\ref{eq35}). 
The main reason to use the local approximation is 
that is impossible to select an effective quadrature
to estimate the 
integral over $l$ due to the specific form of the $\rho(l,z)$ for large $l$'s 
(which can develop a second maxima).

\begin{figure}
\mbox{\includegraphics*[width=2.5in, angle=0]{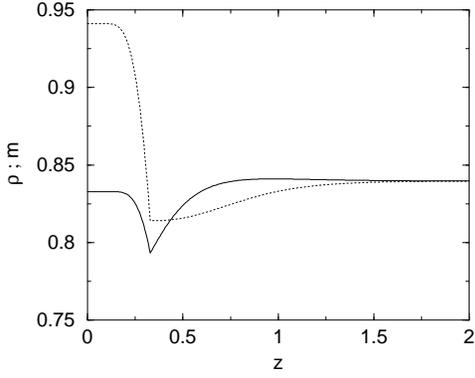}}
\caption[]{Density profile $\rho(z)$ (solid line) and the moment profile
$m(z)$ (dashed line) for $\Delta=0.4298$ and $\rho_0=0.8397$.}
\label{fig23}
\end{figure}

Far from the wetting transition in the region of $\rho_0$'s where the slope of 
$\gamma_{WI}$ with respect to $\rho_0$ is practically zero we have 
calculated the density profiles 
$\rho(z)$ and $m(z)$ for $\Delta=0.4298$. In figure 
\ref{fig23} are given 
these profiles for $\rho_0=0.8397$. As we can see from the figure whereas 
the density adsorption coefficient  $\Gamma^{(0)}$ is negative, the first 
moment adsorption coefficient  
$\Gamma^{(1)}$ is positive. As I have already shown 
(see eq. (\ref{eq47})) the slope of the surface tension depends on certain 
combination of $\Gamma^{(0)}$ and $\Gamma^{(1)}$. For this particular case 
the slope is negative although $\Gamma^{(0)}<0$ in contrast with the result 
obtained using the interface Gibbs-Duhem equation for one component system.

\begin{figure}
\mbox{\includegraphics*[width=3.in, angle=0]{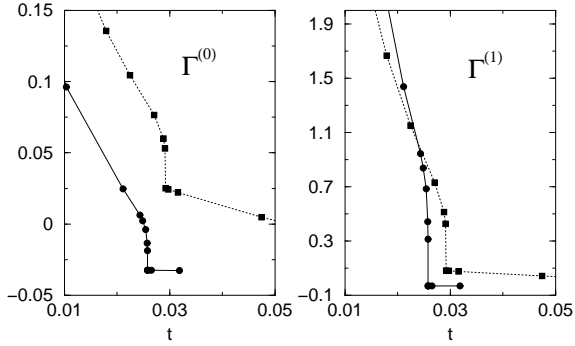}}
\caption[]{Left figure: The density adsorption coefficient $\Gamma^{(0)}$ vs. 
$t=1-\rho_0/\rho_w$ for $\Delta=0.5104$ (solid line) and 
$\Delta=0.4298$ (dashed line). 
Right figure: The first moment adsorption coefficient $\Gamma^{(1)}$ vs. $t$ for the same 
$\Delta$'s.}
\label{fig24}
\end{figure}

For some  $\Delta$'s greater than $\Delta^*$ 
(the critical point value for the 
first order $N_1-N_2$ transition at the wall) 
were calculated 
the adsorption coefficients $\Gamma^{(0)}$ and $\Gamma^{(1)}$ 
as a function of $\rho_0$. 
The results are 
plotted in figure \ref{fig24}. For $\Delta=0.4298$ (the solid line 
in the figure) increasing 
$\rho_0$ occur first a 
second order $I-N_1$ transition 
and then  
a first order $N_1-N_2$ transition in the plane of the wall. 
Was already shown that the 
surface length of the $N_2$  phase increase from zero  
resulting in a second 
order surface transition although in $2D$ is first order. Finally the 
length of the 
$N_2$ surface phase diverge continuously to infinity 
at the complete wetting transition point. 
For $\Delta=0.5104$ (the dotted line in the figure) 
was found a first order $I-N$ transition in the plane of 
the wall and then the complete wetting transition.

\begin{figure}
\mbox{\includegraphics*[width=3.in, angle=0]{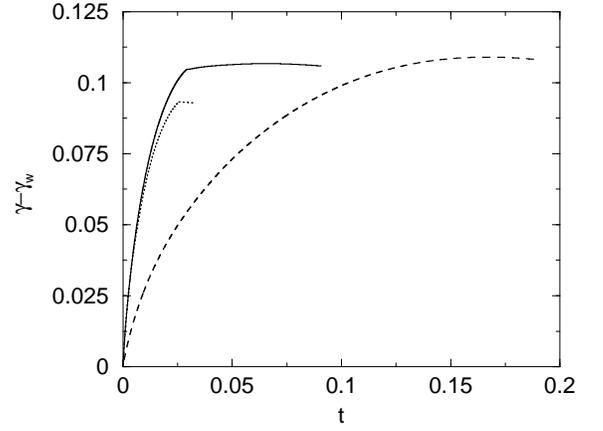}}
\caption[]{The magnitude $\Delta \gamma=\gamma-\gamma_w$ vs. $t$ for 
three different $\Delta$'s: 0.3837 (solid line), 0.4298 
(dashed line) and 0.5104 (dotted line).}
\label{fig25}
\end{figure}

As we can see from this figure
these surface transitions although 
second order are very hard, i.e. there is not a jump 
in the graphic of $\Gamma^{(0,1)}$ as a function of $t$  
but its slopes have a strong discontinuity at the transition points which 
translates in a discontinuity in the second derivative of $\gamma_{WI}$ 
with respect to $t$ as we can see from figure \ref{fig25}. In the 
figure is also plotted the surface tension for $\Delta=0.3837$ corresponding 
to the bulk triple point. For this value the transition to the biaxial nematic 
surface phase is second order (see figure \ref{fig16}) resulting 
in the continuity of $d\Gamma^{(0,1)}/dt$ ($k=0,1$) at this point and then the 
surface tension curvature is also continuous. 

\begin{figure}
\mbox{\includegraphics*[width=2.5in, angle=0]{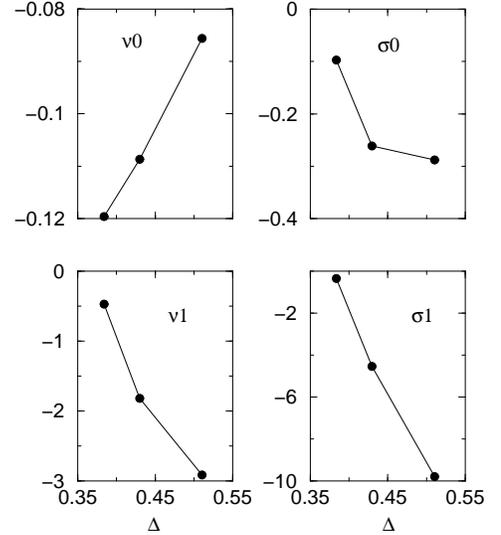}}
\caption[]{The coefficients $\nu_{0,1}$ and $\sigma_{0,1}$ of 
the logarithmic divergence of $\Gamma^{(0,1)}$ 
(see eq (\ref{eq49})). The 
points represent calculated values for $\Delta=0.3837$, 0.4298 and 
0.5104.}
\label{fig26}
\end{figure}

The coefficients $\nu_{0,1}$ and $\sigma_{0,1}$ of the logarithmic 
divergence of $\Gamma^{(0)}$ and $\Gamma^{(1)}$ for log-normal $p(l)$ 
have the same qualitative behaviour (see figure \ref{fig26}) 
as comparing to the Schultz $p(l)$,  
i.e. $\nu_0$ ($\nu_1$) is 
a increasing (decreasing) function of $\Delta$.  
As I have already shown this behaviour is due to the monotonic 
dependence of the differences $\rho^{(N)}-\rho^{(I)}$ (decrease with 
$\Delta$) and $m_N-m_I$ (increase with delta) corresponding 
to the coexistence between shadow nematic and cloud isotropic phases. As 
we can see from figure \ref{fig7} $m^{(N)}$ has a strong maximun 
at $\Delta\sim 0.5$ so is possible that $\nu_1$ becomes 
non monotonic function for higher $\Delta$'s. 

\subsection{Enhancement of capillary nematization by polydispersity}
\label{capillary}
Finally I have studied the capillary nematization of the slit as 
a function of polydispersity. In \cite{Roij1} was studied 
this phenomena for the one-component system. Here I pretend to study 
the role of polydispersity in the possible surface transitions 
governed by confinement. 
The calculations was done in the following 
way. Fixing the distance between two hard walls and fixing 
the density distribution $\rho_0(l)$ of the reservoir (which is isotropic) in
chemical equilibrium with the fluid in the slit, I minimize (\ref{eq24}), this 
time with an external potential which include one more wall 
at $z=h$. For small values of $\rho_0$ the density profile is composed 
approximately by two wall-isotropic 
interfaces, but there is some $\rho_0^*$ (which is 
less than the complete wetting density $\rho_w$ for a single wall-isotropic 
interface) for which the system overcome 
a first order transition to a phase in which the slit is completely 
filled by the nematic phase. This was called capillary nematization 
in analogy with the usual capillary condensation \cite{Roij1}. 
There is some value of the distance between the walls $h^*$ 
(the critical point value) for which the transition becomes second order. 
The natural question that arise is how depends this phenomena with 
polydispersity. Increasing polydispersity the entropic ordering effect that has 
the wall on the fluid particles 
also increase due to the presence of large rods. Then is reasonable to predict 
that the capillary nematization  will be 
enhanced by polydispersity, i.e. fixing  
the distance between the walls the density value $\rho_0$ 
for which occur the transition will decrease 
with polydispersity, and the critical point  
should move to higher $h$'s as was obtained by numerical minimization.

\begin{figure}
\mbox{\includegraphics*[width=2.5in, angle=0]{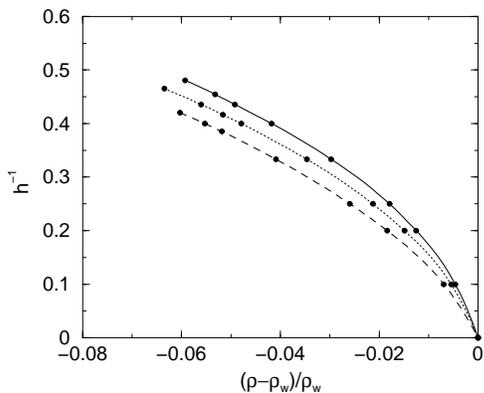}}
\caption[]{The inverse of the width of the slit as a function of the 
reduced density for which occur the capillary nematization. Are 
plotted three curves for different $\Delta$s: 0.064 (solid line), 
0.333 (dotted line) and 0.577 (dashed line).}
\label{fig27}
\end{figure}

The effect that has the polydispersity on capillary nematization is shown  
in figure (\ref{fig27}), where are plotted the 
transition values $h^*$-$t=1-\rho/\rho_w$. For fixed $t$ 
the width $h$ increase with polydispersity.
\section{Conclusions}
\label{conclusions}
Using simple Zwanzig model in the Onsager limit 
for a polydisperse hard rod system I have study how depends the 
phase diagram topology on the selected length distribution  
function $p(l)$. I have found that the relevant property which
affect the bulk phase behaviour is the decaying law 
of $p(l)$ at large $l$'s. 
For distribution functions decaying exponentially or more rapidly  
was found a single coexistence between 
an isotropic and  nematic phase for any polydispersity. 
The cloud isotropic and the shadow 
nematic coexisting densities are decreasing function of $\Delta$, 
whereas the first moments are increasing functions.
For distribution functions decaying slower than an exponential
always we can find a maximun extreme value for the lengths of particles 
($L_{\rm{max}}$) for which the 
bulk phase diagram turns out to be more complex. Now there is some $\Delta$ 
for which the cloud isotropic and shadow nematic curves 
($x_N=0$) develop a triple 
point: one isotropic ($I_1$) coexist with two different 
nematics ($N_1$ and $N_2$), one of them ($N_2$) with a bimodal 
distribution function. The region of triple coexistence extend until 
very small values of $x_N$ which strongly depend on the 
decaying law  and on the maximun extreme value
$L_{\rm{max}}$.
For values of $x_N$ and $\Delta$ outside the triple coexistence region 
the shadow nematic curve becomes continuous keeping its usual behaviour 
(a decreasing function of $\Delta$).
Extending in the appropriate way the interface Gibbs-Duhem equation for the 
thermodynamic of interfaces  
to the polydisperse system I have found that the slope of the surface 
tension depends on certain combination of different moment adsorption 
coefficients which was confirmed by numerical minimization. 
I have studied also the surface phase diagram of the hard rod system 
interacting with a hard wall for two representative 
families of distribution functions. Each of them produce 
two qualitatively different surface phase diagrams. With the particular choice 
of the Schultz's distribution function from the first family were calculated 
the equilibrium profiles and different adsorption coefficients. 
Increasing the bulk density value was found some $\rho^*_0$ for which  
the system overcomes  a second order uniaxial-biaxial nematic transition 
The thickness 
of the new biaxial nematic layer increase from zero and diverges continuously 
at the wetting transition density $\rho_w$. Both $\rho^*_0$ and 
$\rho_w$ are decreasing functions of polydispersity.  
For the log-normal distribution function (a particular choice from 
the second family) the surface phase diagram changes dramatically. 
The second order $I-N_1$ transition at the wall is followed by 
a two dimensional first order transition between two nematics 
($N_1$ and $N_2$). The thickness 
of the $N_2$ layer increase from zero continuously and diverge at the wetting 
density. Increasing $\Delta$ this $N_1-N_2$ transition coalesce with  
the second order $I-N_1$ transition and for higher values we find 
a single first order $I-N_2$ at the wall followed by the complete 
wetting. 
For the log-normal distribution 
function I have obtained using the local approximation 
that although very hard (the slopes of the adsorption coefficients 
at the transition points 
have strong discontinuities) these transitions 
are second order in nature. 
Was studied also the segregation phenomena occurring at the isotropic-nematic
interfaces. The numerical calculations show a preferential 
segregation of long rods to the nematic side of the interface being  
very sensitive this phenomena to the degree of polydispersity.
Finally was shown that polydispersity enhance the capillary nematization 
of the slit.  
\begin{acknowledgements}
It is a pleasure to thank Jos\'e A. Cuesta and Carlos Rasc\'on 
for stimulating discussions. 
Is gratefully acknowledged the support from 
the postdoctoral position at Wuppertal University and from the Direcci\'on General de Investigaci\'on, Conserjer\'{\i}a 
de Educaci\'on de la Comunidad de Madrid.  
This work is part of Project 
No. BFM2000-0004 of the Direcci\'on General de Ense\~nanza Superior. 
\end{acknowledgements}

\end{document}